\documentclass[traditabstract]{aa}

\def\setsymbol#1#2{\expandafter\def\csname #1\endcsname{#2}}
\def\getsymbol#1{\csname #1\endcsname}

\def\Planck{\textit{Planck}}

\newbox\tablebox    \newdimen\tablewidth
\def\leaderfil{\leaders\hbox to 5pt{\hss.\hss}\hfil}
\def\tablenote#1 #2\par{\begingroup \parindent=0.8em
    \abovedisplayshortskip=0pt\belowdisplayshortskip=0pt
    \noindent
    $$\hss\vbox{\hsize\tablewidth \hangindent=\parindent \hangafter=1 \noindent
    \hbox to \parindent{$^#1$\hss}\strut#2\strut\par}\hss$$
    \endgroup}

\def\L2{\ifmmode L_2\else $L_2$\fi}

\def\DeltaT{\ifmmode \Delta T\else $\Delta T$\fi}
\def\deltat{\ifmmode \Delta t\else $\Delta t$\fi}
\def\fknee{\ifmmode f_{\rm knee}\else $f_{\rm knee}$\fi}
\def\Fmax{\ifmmode F_{\rm max}\else $F_{\rm max}$\fi}
\def\solar{\ifmmode{\rm M}_{\mathord\odot}\else${\rm M}_{\mathord\odot}$\fi}
\def\Msolar{\ifmmode{\rm M}_{\mathord\odot}\else${\rm M}_{\mathord\odot}$\fi}
\def\Lsolar{\ifmmode{\rm L}_{\mathord\odot}\else${\rm L}_{\mathord\odot}$\fi}
\def\inv{\ifmmode^{-1}\else$^{-1}$\fi}
\def\mo{\ifmmode^{-1}\else$^{-1}$\fi}
\def\sup#1{\ifmmode ^{\rm #1}\else $^{\rm #1}$\fi}
\def\expo#1{\ifmmode \times 10^{#1}\else $\times 10^{#1}$\fi}
\def\,{\thinspace}
\def\lsim{\mathrel{\raise .4ex\hbox{\rlap{$<$}\lower 1.2ex\hbox{$\sim$}}}}
\def\gsim{\mathrel{\raise .4ex\hbox{\rlap{$>$}\lower 1.2ex\hbox{$\sim$}}}}

\def\simprop{\mathrel{\raise .4ex\hbox{\rlap{$\propto$}\lower 1.2ex\hbox{$\sim$}}}}
\def\deg{\ifmmode^\circ\else$^\circ$\fi}
\def\pdeg{\ifmmode $\setbox0=\hbox{$^{\circ}$}\rlap{\hskip.11\wd0 .}$^{\circ}
          \else \setbox0=\hbox{$^{\circ}$}\rlap{\hskip.11\wd0 .}$^{\circ}$\fi}
\def\arcs{\ifmmode {^{\scriptstyle\prime\prime}}
          \else $^{\scriptstyle\prime\prime}$\fi}
\def\arcm{\ifmmode {^{\scriptstyle\prime}}
          \else $^{\scriptstyle\prime}$\fi}
\newdimen\sa  \newdimen\sb
\def\parcs{\sa=.07em \sb=.03em
     \ifmmode \hbox{\rlap{.}}^{\scriptstyle\prime\kern -\sb\prime}\hbox{\kern -\sa}
     \else \rlap{.}$^{\scriptstyle\prime\kern -\sb\prime}$\kern -\sa\fi}
\def\parcm{\sa=.08em \sb=.03em
     \ifmmode \hbox{\rlap{.}\kern\sa}^{\scriptstyle\prime}\hbox{\kern-\sb}
     \else \rlap{.}\kern\sa$^{\scriptstyle\prime}$\kern-\sb\fi}
\def\ra[#1 #2 #3.#4]{#1\sup{h}#2\sup{m}#3\sup{s}\llap.#4}
\def\dec[#1 #2 #3.#4]{#1\deg#2\arcm#3\arcs\llap.#4}
\def\deco[#1 #2 #3]{#1\deg#2\arcm#3\arcs}
\def\rra[#1 #2]{#1\sup{h}#2\sup{m}}

\def\dots{\relax\ifmmode \ldots\else $\ldots$\fi}
\def\WHzsr{\ifmmode $W\,Hz\mo\,sr\mo$\else W\,Hz\mo\,sr\mo\fi}
\def\mHz{\ifmmode $\,mHz$\else \,mHz\fi}
\def\GHz{\ifmmode $\,GHz$\else \,GHz\fi}
\def\mKs{\ifmmode $\,mK\,s$^{1/2}\else \,mK\,s$^{1/2}$\fi}
\def\muKs{\ifmmode \,\mu$K\,s$^{1/2}\else \,$\mu$K\,s$^{1/2}$\fi}
\def\muKRJs{\ifmmode \,\mu$K$_{\rm RJ}$\,s$^{1/2}\else \,$\mu$K$_{\rm RJ}$\,s$^{1/2}$\fi}
\def\muKHz{\ifmmode \,\mu$K\,Hz$^{-1/2}\else \,$\mu$K\,Hz$^{-1/2}$\fi}
\def\MJysr{\ifmmode \,$MJy\,sr\mo$\else \,MJy\,sr\mo\fi}
\def\MJysrmK{\ifmmode \,$MJy\,sr\mo$\,mK$_{\rm CMB}\mo\else \,MJy\,sr\mo\,mK$_{\rm CMB}\mo$\fi}
\def\microns{\ifmmode \,\mu$m$\else \,$\mu$m\fi}

\def\muK{\ifmmode \,\mu$K$\else \,$\mu$\hbox{K}\fi}
\def\microK{\ifmmode \,\mu$K$\else \,$\mu$\hbox{K}\fi}
\def\muW{\ifmmode \,\mu$W$\else \,$\mu$\hbox{W}\fi}
\def\kms{\ifmmode $\,km\,s$^{-1}\else \,km\,s$^{-1}$\fi}
\def\kmsMpc{\ifmmode $\,\kms\,Mpc\mo$\else \,\kms\,Mpc\mo\fi}
\providecommand{\sorthelp}[1]{}

\usepackage{colortbl}
\usepackage[nonamebreak]{natbib}
\usepackage[stable]{footmisc}
\usepackage{amsmath}
\usepackage{txfonts}
\usepackage{placeins}
\usepackage{appendix}
\bibpunct{(}{)}{;}{a}{}{,}
\usepackage{graphicx}
\usepackage{epstopdf}
\usepackage{ifthen} 
\usepackage[breaklinks, colorlinks, citecolor=blue]{hyperref}
\usepackage{overpic}
\usepackage{fixltx2e}
\usepackage{rotating}
\usepackage{subfigure}
\usepackage{epsfig}
\usepackage{multirow}
\usepackage{array}
\usepackage{color}
\usepackage[usernames,dvipsnames]{xcolor}
\usepackage{psfrag}
\usepackage[T1]{fontenc}
\usepackage{url}
\usepackage{xfrac}
\usepackage{subeqnarray}
\usepackage{breakurl}

\usepackage{xcolor} 
\usepackage{makecell} %
\usepackage{soul}

\newcolumntype{L}[1]{>{\raggedright\let\newline\\\arraybackslash\hspace{0pt}}m{#1}}
\newcolumntype{C}[1]{>{\centering\let\newline\\\arraybackslash\hspace{0pt}}m{#1}}
\newcolumntype{R}[1]{>{\raggedleft\let\newline\\\arraybackslash\hspace{0pt}}m{#1}}

\definecolor{mywarningcolor}{RGB}{208,59,32} 
\newcommand{\warning}[1]{} 
\begin{document}

\topmargin=-1cm
\oddsidemargin=-1cm
\evensidemargin=-1cm
\textwidth=17cm
\textheight=25cm
\raggedbottom
\sloppy

\definecolor{Blue}{rgb}{0.,0.,1.}
\definecolor{LightSkyBlue}{rgb}{0.691,0.827,1.}
\definecolor{Red}{rgb}{1.,0.,0.}
\definecolor{Green}{rgb}{0.,1.,0.}
\definecolor{Try}{rgb}{0.15,0.,1}
\definecolor{Black}{rgb}{0., 0., 0.}

\def\fsky{$f_{\rm sky}$}
\def\microK{\,$\mu$K}
\def\microKCMB{\,$\mu$K$_{\textit{CMB}}$}
\def\ellmax{$\ell_{max}$}
\def\deltabetadust{\Delta\beta\textit{Dust}}
\def\deltatdust{\Delta\textit{TDust}}
\def\Nside{\ensuremath{N_{\mathrm{side}}}}

\newcommand{\citelowell}{\citetalias{planck2014-a10}}
\newcommand{\citedpc}{\citetalias{planck2016-l03}}
\newcommand{\citesrolltwo}{\citetalias{2019A&A...629A..38D}}

\defcitealias{planck2014-a10}{LowEll paper}
\defcitealias{planck2016-l03}{L03 paper}
\defcitealias{2019A&A...629A..38D}{SRoll2 paper}

\newcommand{\sroll}{{\tt SRoll}}
\newcommand{\srollone}{{\tt SRoll1}}
\newcommand{\srolltwo}{{\tt SRoll2}}
\newcommand{\Bcsep}{{\tt Bcsep}}
\newcommand{\bmodes}{{\it B}-modes}
\newcommand{\emodes}{{\it E}-modes}
\newcommand{\hmone}{{\it half}-mission1}
\newcommand{\hmtwo}{{\it half}-mission2}
\newcommand{\full}{{\it FULL}}

\newcommand{\jlpcomment}[1]{{\bf \color{red} (#1)}}
\newcommand{\arcomment}[1]{{\bf \color{blue} (#1)}}
\newcommand{\jacomment}[1]{{\bf \color{orange} (#1)}}
\newcommand{\jmcomment}[1]{{\bf \color{green} (#1)}}
\newcommand{\fbcomment}[1]{{\bf \color{purple} (#1)}}
\newcommand{\lvcomment}[1]{{\bf \color{violet} #1}}
\newcommand{\jacorrect}[2]{{\st{#1} \bf \color{orange} #2}}
\newcommand{\lvcorrect}[2]{{\st{#1} \bf \color{violet} #2}}
\newcommand{\vgcomment}[1]{{\bf \color{cyan}[VG:#1]}}
\newcommand{\vgcorrect}[2]{{\st{#1} \bf \color{cyan} #2}}

\title{Dust polarization spectral dependence from Planck HFI data}
\subtitle{Turning point on CMB polarization foregrounds modelling}

\author{
Alessia~Ritacco\inst{1,2,3}~\thanks{Corresponding author: A.~Ritacco, alessia.ritacco@inaf.it}
\and
Fran\c{c}ois~Boulanger\inst{2}
\and
Vincent~Guillet\inst{3,4}
\and
Jean-Marc~Delouis\inst{5}
\and
Jean-Loup~Puget\inst{2,3}
\and
Jonathan~Aumont\inst{6}
\and
L\'eo~Vacher\inst{6}}

\institute{
INAF-Osservatorio Astronomico di Cagliari, Via della Scienza 5, 09047 Selargius, IT
\and
Laboratoire de Physique de l’$\acute{\rm E}$cole Normale Sup$\acute{\rm e}$rieure, ENS, Universit$\acute{\rm e}$ PSL, CNRS, Sorbonne Universit$\acute{\rm e}$, Universit$\acute{\rm e}$ de Paris, 75005 Paris, France
\and
Institut d'Astrophysique Spatiale, CNRS, Universit\'{e} Paris-Saclay, CNRS, B\^{a}t. 121, 91405 Orsay, France\goodbreak
\and
Laboratoire Univers et Particules de Montpellier, Universit\'e de
Montpellier, CNRS/IN2P3, CC 72, Place Eug\'ene Bataillon, 34095
Montpellier Cedex 5, France
\and
Laboratoire d'Oc{\'e}anographie Physique et Spatiale (LOPS), Univ. Brest, CNRS, Ifremer, IRD, Brest, France\goodbreak
\and
IRAP, Universit$\acute{\rm e}$ de Toulouse, CNRS, CNES, UPS, (Toulouse), France
}
\abstract{The search for the primordial \bmodes\ of the cosmic microwave background (CMB) relies on the separation  from the brighter foreground dust signal. 
In this context, the characterisation of the spectral energy distribution (SED) of thermal dust in polarization has  become a critical subject of study.
We present a power-spectra analysis of \Planck\ data, which improves on previous studies by using the newly released \srolltwo\ maps that correct residual data systematics, and by extending the analysis to regions near the Galactic plane. Our analysis focuses  on the lowest multipoles between $\ell$=4 and 32, and three sky areas with sky fractions of $f_{\rm sky} = 80$\%, 90\%, and 97\%. 
The mean dust SED for polarization and the 353\,GHz $Q$ and $U$ maps are used to compute residual maps at 100, 143 and 217\,GHz, highlighting variations of the dust polarization SED on the sky and along the line of sight. Residuals are detected at the three frequencies for the three sky areas. 
We show that models based on total intensity data are underestimating by a significant factor the complexity of dust polarized CMB foreground. Our analysis emphasizes the need to include variations of polarization angles of the dust polarized CMB foreground. 
The frequency dependence of the $EE$ and $BB$ power spectra of the residual maps yields further insight. We find that the moments expansion to the first order of the modified black-body (MBB) spectrum provides a good fit to the $EE$ power-spectra. This result suggests that the residuals could follow mainly from variations of dust MBB spectral parameters. 
However, this conclusion is challenged by cross-spectra showing that the residuals maps at the three frequencies are not fully correlated, and the fact that the $BB$ power-spectra do not match the first order moment expansion of a MBB SED. 
This work sets new requirements for simulations of the dust polarized foreground and component separation methods, showing that a significant refinement to dust modelling is necessary to ensure an unbiased detection of the CMB primordial \bmodes~at the precision required by future CMB experiments. Further work is also needed to model theoretically the impact of polarization angle variations on $EE$ and $BB$ power spectra of residuals maps.

}

\maketitle

\section{Introduction}
One of the outstanding questions in cosmology concerns the existence of primordial gravitational waves as predicted by the theory of cosmic inflation \citep{Guth1981, LINDE1982}. Although the primordial gravitational waves are not directly detectable with foreseen experiments, they are expected to leave an imprint as a curl-like pattern in the cosmic microwave background (CMB) polarization anisotropies, referred to as primordial \bmodes, which could be measured.
This is a main goal of present and future CMB experiments, including the  {\it LiteBIRD} satellite \citep{litebird2022}, and {\it BICEP/Keck} \citep{Ade_BICEP_22}, the {\it Simons Observatory} \citep{Simons19} and {\it CMB-Stage~4} \citep{cmbs4} from the ground. These experiments aim at measuring the tensor-to-scalar ratio $r$, which represents the amplitude of the primordial tensor (\bmodes) relative to scalar perturbations (\emodes) of the CMB. This parameter, related to  the energy scale of inflation, is expected to be in the range between 10$^{-2}$ to 10$^{-4}$ \citep{Kamionkowski16,litebird2022}. A very high control and subtraction of instrumental systematic effects and Galactic foregrounds is required for an unbiased measurement of such low $r$ values.

The dust emission represents a major obstacle, because its amplitude is much larger than that of primordial \bmodes\ signal  \citep{planck_XI2020}. In this context, the characterisation of the spectral energy distribution (SED) of dust has become a critical step in the search for \bmodes. 
The \Planck\ data have shown that the mean dust SED for polarization and total intensity are very close, and that they are both well fitted by a modified black body (MBB) law  \citep{2015A&A...576A.107P,planck_XI2020}. 
In addition, far-IR polarization measurements obtained by the {\it BLASTPol} balloon-borne experiment \citep[e.g.,][]{Ashton18}, have shown that the dust polarization fraction is roughly constant  between $250\,\mu$m and 3\,mm~(100\,GHz). These remarkable results suggest that the emission from a single grain type dominates the long-wavelength emission in both polarization
and total intensity \citep{Guillet2018,Hensley22}.

Maps of MBB parameters (dust spectral index and temperature)
have been obtained fitting the total intensity \Planck\ data \citep[e.g.][]{2016A&A...594A..10P,2016A&A...596A.109P}. 
The \Planck\ 2018 data release \citep[hereafter PR3,][]{planck2016-l03} does not provide comparable constraints on MBB parameters in polarization due to insufficient signal to noise ratio and instrumental systematics \citep{Osumi21}. 
These limiting factors are emphasized by the lack of polarization maps  at 545 and 857 GHz to constrain the dust SED.
The frequency dependence of dust polarization data also involves polarization angles. Where dust SED and magnetic field orientation vary within the beam, the frequency scaling of the Stokes $Q$ and $U$ parameters may differ \citep{Ichiki19,Vacher22}. Integration along the line of sight may therefore induce variations of the polarization angle with frequency \citep{tassis2014,PIPL}. \citet{Pelgrims21} provided first observational evidence  of this effect analyzing \Planck\ data toward lines of sight with multiple velocity components in H~I emission. Additional emission components (e.g. magnetic dipole and CO emission \citep{puglisi2017}) can also contribute to SED variations \citep{Hensley_2018}.  

SED variations on the sky and along the line of sight induce a decorrelation between dust
emission at different frequencies, which is referred to as frequency decorrelation. Attempts to detect frequency decorrelation through a power spectra analysis of the multi-frequency \Planck\ data have only yielded upper limits \citep{PIPL,planck_XI2020}. 
This work improves on previous studies by using a new upgraded version of \Planck\ maps.  
The polarization maps at frequencies 100 to 353\,GHz used in this paper are taken from the \srolltwo.0\ version of the $Planck$ data processing \citep{delouis2018}, which correct the data from  systematics that plague 
the PR3 release. This work also improves the sensitivity to frequency decorrelation  by extending the analysis from the high Galactic latitude sky regions best suited for CMB observations 
to brighter regions near the Galactic plane. 



The paper is organized as follows. Sect.~\ref{sec:planck_data} presents the \Planck\ data we use; Sect.~\ref{sec:reference_model} introduces the  reference models we use for data simulations and analysis. 
In Sect.~\ref{sec: dust SED}, we determine the dust mean SED for polarization from 100 to 353 GHz.
We quantify spatial variations of the dust polarization SED in Sect.~\ref{sec:spatial_variations} and the contribution of polarization angles in  Sect.~\ref{sec:dust_polarization_angle}.  
The frequency dependence is analyzed in Sect.~\ref{sec:moments_expansion}.
The paper results are summarized in Sect.~\ref{sec:conclusions}. 

\section{{\it \bf Planck} polarization data and masks}
\label{sec:planck_data}
The $Planck$ satellite  observed the sky in total intensity (also referred as temperature) in the range of frequency of the electromagnetic spectrum from 30 to 857\,GHz, and in polarization from 30 to 353\,GHz. 
Data was obtained from two instruments on board the satellite: the Low Frequency Instrument \citep[LFI,][]{2011A&A...536A...3M}, and the High Frequency Instrument \citep[HFI,][]{planck2011}. 
$Planck$ HFI measured the linear polarization at 100, 143, 217, and 353\,GHz \citep{rosset}. 

The \Planck~scanning strategy sampled almost all the sky pixels every six months, with alternating scan directions in successive six-month periods. The \Planck~mission includes five surveys, each covering a large fraction of the sky (hereafter $f_{\rm sky}$). Maps are produced for the full-mission
data-set together with the survey, year, and half-mission maps, as reported in \cite{planck2015viii}.

\subsection{\srolltwo\ maps}
In this work, we use sky maps and end-to-end data simulations produced by the \srolltwo~software\footnote{The \srolltwo~maps are available \href{http://sroll20.ias.u-psud.fr/sroll20_data.html}{here}~and the simulations \href{http://sroll20.ias.u-psud.fr/sroll20_sim.html}{here}
.}. This latter has been developed to improve subtraction of systematic effects. 
The dominant systematic effect for the polarized signal at 353\,GHz in the PR3 maps is related to the poor measurement of the time transfer function of the detectors, while at lower frequencies it is dominated by the non-linearity of the analog-to-digital converters.
Both systematics have been greatly improved consistently with all other known effects for the \srolltwo\ data-set \citep{delouis2018}.

Hereafter $Q_{\rm P}(\nu)$ and $U_{\rm P}(\nu)$ are the $Planck$ polarization maps at frequency $\nu$, including the CMB, dust and synchrotron as well as noise and systematics.
We use the \Planck~HFI maps at 100, 143, 217 and $353\,$GHz. We limit our data analysis to polarization at low multipoles (4<$\ell$<$\ell_{\rm max}$=32) and work with HEALPix pixelization \citep{healpix} at \Nside=32 (i.e. map pixel size of $1.8^{\circ}$). In order to obtain these maps, we follow  \citet{2016A&A...596A.107P} and \citet{2020A&A...641A...5P} degrading the full-resolution maps first to \Nside=1024, to ease the computation, and next to \Nside=32 applying the following cosine filter in harmonic space \citep{2009MNRAS.400..219B}:
\begin{equation}
f(\ell)=\left \{\begin{array}{ll}
1,\qquad \qquad \qquad \qquad \quad \ell\leqslant \Nside;\\
\frac{1}{2}\left(1+\sin\left(\frac{\pi}{2}\frac{\ell}{\Nside}\right)\right),\quad \Nside < \ell < 3\Nside;\\
0,\qquad \qquad \qquad \qquad \quad \ell\geqslant 3\Nside.
\end{array}
\right.
\end{equation}
All the maps used in this study and presented in the paper have been degraded to \Nside=32 by following this approach. 
\begin{figure*}[h!]
    \centering
    \includegraphics[width=0.95\textwidth]{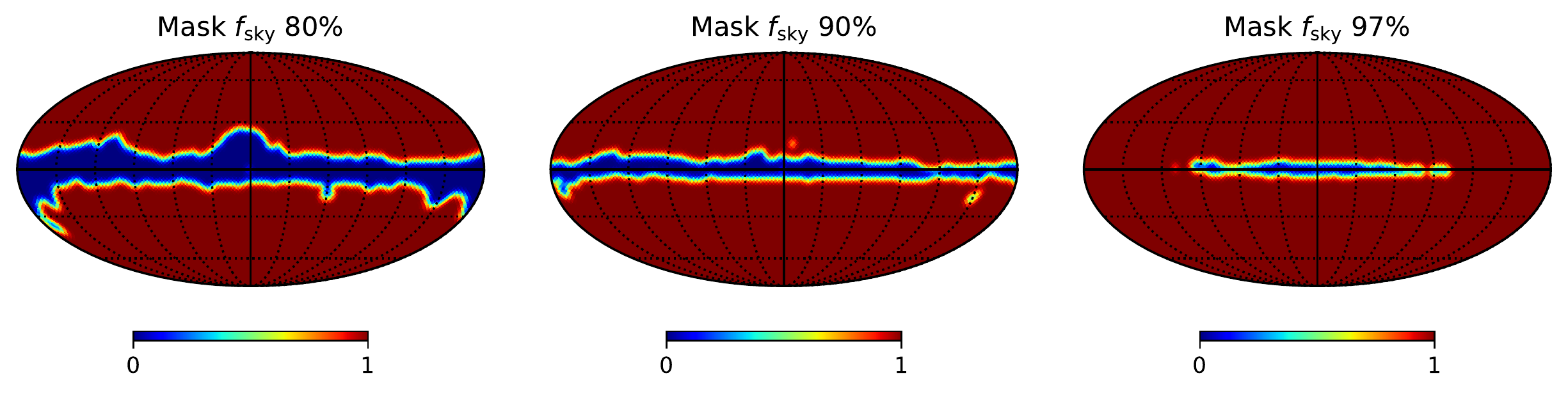}
    \caption{From left to right apodized masks for $f_{\rm sky}$=80, 90, and 97\,\%.}
    \label{fig:masks}
\end{figure*}

\subsection{Subtraction of synchrotron emission}
\label{sec:synchrotron_subtraction}
The 100 and 143\,GHz maps include non-negligible synchrotron emission that we subtract to focus our data analysis on the dust emission. 
Hence, we define the following maps
\begin{eqnarray}
    Q^{\prime}_{\rm P}(\nu) &= Q_{\rm P}(\nu) - Q_{\rm s}(\nu) \nonumber \\
    U^{\prime}_{\rm P}(\nu) &= U_{\rm P}(\nu) - U_{\rm s}(\nu),
    \label{eq:dust_pure}
\end{eqnarray}
where $[Q_{\rm s},U_{\rm s}](\nu)$ are estimates of the synchrotron Stokes parameters. Notice that the $Q^{\prime}_{\rm P}(\nu)$ and $U^{\prime}_{\rm P}(\nu)$ maps include the CMB. Hereafter the prime superscript is used to indicate Stokes maps where the synchrotron emission is subtracted or absent.

We use synchrotron template maps at $\nu_{\rm s}$ = 30\,GHz, as obtained by the {\it Commander} component separation \citep{2016A&A...594A..10P} applied to  PR3 $Planck$ maps. To extrapolate from 30 to 100 and 143\,GHz, we use a single spectral index $\beta_{\rm s}$, uniform over the sky, to extrapolate synchrotron polarization from 30 to $\nu=[100,143]$ GHz.
\begin{eqnarray}
\begin{aligned}
&[Q,U]_{\rm s}(\nu)=[Q,U]_{\rm s}(\nu_{\rm s}) \cdot  C^{\rm CC_s}_{\nu} \cdot  {C_{\nu}^{UC_{\rm RJ}}} \cdot\left(\frac{\nu}{30\,{\rm GHz}}\right)^{\beta_{\rm s}},
\label{eq:synchrotron}
\end{aligned}
\end{eqnarray}
where the $[Q,U]_{\rm s}(\nu)$ maps are in $K_{\rm CMB}$ and $[Q,U]_{\rm s}(\nu_{\rm s})$ in $K_{\rm RJ}$; 
C$_{\nu}^{UC_{\rm RJ}}$ is the conversion factor from $K_{\rm RJ}$ to $K_{\rm CMB}$, and $C^{\rm CC_s}_{\nu}$  the color correction, at frequency $\nu$. 
The conversion factor values at 100 and 143\,GHz are listed in Tab.~\ref{tab:units}.

To determine $\beta_{\rm s}$, we use spectral indices derived by \cite{2022JCAP...04..003M}, from a detailed analysis combining $Planck$ and $WMAP$ \citep{2013ApJS..208...20B} data at 30 and 23\,GHz, respectively. 
Table~4 in \cite{2022JCAP...04..003M} lists spectral indices derived from $EE$ and $BB$ power spectra. We combine the two pairs of $EE$ and $BB$ spectral indices, for the two largest sky areas with $f_{\rm sky} = 94$ and 70\,\%, to compute a mean value, weighted by inverse squared uncertainties, $\beta_{\rm s} = -3.19 \pm 0.07$.

This mean value is a reasonable approximation considering recent studies that obtained a synchrotron spectral index in polarization $\beta_{s}$ of: i) $-3.22 \pm 0.08$ over the southern sky \citep{Nicoletta2018}; ii) $-3.17\pm0.06$ in the North Polar Spur \citep{2020arXiv201108503S}; and iii) $-3.25 \pm 0.06$ within the BICEP2/Keck survey footprint \citep{Weiland_2022}.
The study of \cite{delahoz2022} hints at evidence of spatial variations of $\beta_{s}$, which would need to be accounted for in future developments. 
 
\begin{table}[h!]
\centering
\arrayrulecolor[rgb]{0.753,0.753,0.753}
\begin{tabular}{!{\color{black}\vrule}l!{\color{black}\vrule}l|l|l|l!{\color{black}\vrule}l|l|l!{\color{black}\vrule}l|l|l!{\color{black}\vrule}}
\arrayrulecolor{black}\cline{2-4}
\arrayrulecolor{black}\hline
$\nu$ [GHz] & 100 & 143 & 217 & 353 \\
\arrayrulecolor{black}\hline
$C^{\rm CC_d}_{\nu}$ & 1.09 & 1.02 & 1.12 & 1.11 \\
\arrayrulecolor[rgb]{0.753,0.753,0.753}\hline
$C_{\nu}^{UC_{K}}$ & 244.1 & 371.7 & 483.7 & 287.4\\
\arrayrulecolor[rgb]{0.753,0.753,0.753}\hline
$C_{\nu}^{UC_{RJ}}$ & 1.26 & 1.69 & &\\
\arrayrulecolor[rgb]{0.753,0.753,0.753}\hline
$C^{\rm CC_s}_{\nu}$ & 0.9797 & 0.9504 & & \\
\arrayrulecolor[rgb]{0.753,0.753,0.753}\hline
$\rho_{\nu}$ & 1.005 & 0.98 & 1.015 & 1 \\

\arrayrulecolor{black}\hline
\end{tabular}
\caption{Unit conversions $C_{\nu}^{UC_{RJ},UC_{K}}$ and colour corrections $C^{\rm CC_{d,s}}_{\nu}$ as taken from \cite{planck_convers_units}. The factor $\rho(\nu)$ is the correction to the polarization efficiencies \citep{planck2016-l03,planck_XI2020}.}
\label{tab:units}
\end{table}

\subsection{Sky masks and power spectra}

We compute power spectra for three sky areas presented in Fig.~\ref{fig:masks} with sky fractions $f_{\rm sky}$ of 80\,\%, 90\,\% and 97\,\%. We use larger $f_{\rm sky}$ than what is usually done to study Galactic foregrounds in order to increase the signal-to-noise ratio. 
To mask areas of bright dust emission, we use the dust optical depth map estimated at 353 GHz by fitting a modified blackbody (MBB) spectral model to the GNILC (generalized needlet internal linear combination \citep[GNILC,][]{GNILC}) dust maps at 353, 545, 857, and the IRAS 3000 GHz map \citep{Fixsen_1999}.
We smooth the map with a 5$^{\circ}$ beam before reducing the HEALPix resolution to \Nside=32. Next, we sort the pixels by increasing amplitude
to define the appropriate masks to obtain $f_{\rm sky} =$ 80\,\%, 90\,\%, and 97\,\%. 
Finally, we smooth again the mask maps to 5$^{\circ}$ beam resolution, in order to apodize them and avoid edge effects on the Galactic cut contours.

The power spectra are computed by using the \href{http://www2.iap.fr/users/hivon/software/PolSpice/}{\it PolSpice} estimator, which corrects for multipole-to-multipole coupling and for the mixing of the $E$- and $B$- modes due to the sky masking  \citep{Chon04}.  
We systematically compute cross-power spectra to have no bias from data noise. To compute power spectra at one given frequency, 
we use the so-called “half-mission” maps
(hereafter “HM”). Notice that most of the residual instrumental systematic effects evolve with time and are decorrelated between the two half-mission data sets. Throughout the paper, we use $\mathcal{D}_\ell \equiv \ell (\ell+1) \, \mathcal{C}_\ell / 2 \pi$ where $\mathcal{C}_\ell $ is the original angular power spectrum. 

\section{Reference models and data simulations}
\label{sec:reference_model}

In our analysis of the $Planck$ data, we make use of two reference
models of the dust emission and of data simulations,
which we introduce in this section.

\subsection{Reference models}
\label{subsec:ref_models}
The $Planck$ data analysis has shown that the MBB emission law fits well the SED of the dust emission for total intensity \citep{2014A&A...571A..11P,2016A&A...594A..10P} and polarization \citep{planck_XI2020}. 
This provides a convenient and commonly used parametrization of the dust SED: 
\begin{eqnarray}
I_{\rm d}(\nu) = C^{\rm UC_K}_\nu \cdot C^{\rm CC_d}_\nu(\beta_{\rm d},T_{\rm d}) \cdot \rho_\nu \cdot \tau_{\nu_0} \left(\frac{\nu}{\nu_{0}}\right)^{\beta_{\rm d}} \cdot B_\nu(T_{\rm d})
  \label{I_nu}
  \end{eqnarray}
where $B_\nu(T_{\rm d})$ is the Planck function; $T_{\rm d}$, $\beta_{\rm d}$ and $\tau_{\nu_0}$ are the dust temperature, spectral index and optical depth maps at the reference frequency $\nu_0$, respectively. 
The MBB emission is expressed in MJy\,sr$^{-1}$ whereas the data are in thermodynamic units. The conversion
between the two is accomplished by two factors. The first, $C^{\rm UC_K}_\nu$, is a
unit conversion from MJy\,sr$^{-1}$ to K$_{\rm CMB}$ for the reference spectral dependence: constant product $\nu  I_\nu$ over the bandpass. The
second, $C^{\rm CC_d}_\nu(\beta_{\rm d},T_{\rm d})$, is the colour correction accounting for the difference between the reference spectral dependence and the MBB spectrum. This correction depends on the MBB parameters: $\beta_{\rm d}$ and $T_{\rm d}$, which are taken equal to 1.53 and 19.6 K, respectively \citep{planck_XI2020}. The estimated values are presented in Tab.~\ref{tab:units}.

In \cite{planck2016-l03}, the polarization efficiencies have been adjusted, within the uncertainties of the  ground calibration, in order to match cosmological parameters derived from CMB polarization with the ones obtained from CMB temperature \citep{planck2016-l03}. 
The factor $\rho_\nu$ in Eq.~\ref{I_nu} represents this correction to the polarization efficiencies. We use the same values as \citet{planck_XI2020}, which are listed in Tab.~\ref{tab:units}. The uncertainty on $\rho_\nu$ is estimated to be 0.5\% at $\nu$=[100, 143, 217]\,GHz. The correction could not be estimated with the required accuracy at 353\,GHz.

We apply Eq.~\ref{I_nu} to two sets of MBB parameters derived from $Planck$ component separations methods: i) the GNILC method using maps corrected for anisotropies of the cosmic infrared background, and ii) the standard Bayesian analysis framework, implemented in the {\it Commander} code \citep{2016A&A...594A..10P}. 
For GNILC, $T_{\rm d}$, $\beta_{\rm d}$ and $\tau_{\nu_0}$ have been obtained by fitting a MBB model on the dust total intensity at the $Planck$ frequencies 353, 545 and $857\,$GHz and IRAS $3000\,$GHz, while the {\it Commander} fit includes all \Planck\ frequencies, together with the 9-year WMAP observations between 23 and 94 GHz \citep{2013ApJS..208...20B} and a 408 MHz survey map \citep{1982A&AS...47....1H}.

To build our reference models, we assume that the dust SED is the same for total intensity and polarization, an hypothesis supported by the close match between the total intensity and polarization SEDs  \citep{2015A&A...576A.107P,planck_XI2020}, 
which we aim to further test in this paper. For GNILC, we also assume that the MBB fitted over the far-IR may be extrapolated to microwave frequencies \citep{2014A&A...566A..55P}.
Within this framework, the Stokes parameters $Q_{\rm d}(\nu)$ and $U_{\rm d}(\nu)$ at $\nu=[100,143,217]$\,GHz may be computed from the $Planck$ maps at $\nu_0=353\,$GHz:
\begin{eqnarray}
\centering
    Q_{\rm d}(\nu) &=& \frac{I_{\rm d}(\nu)}{I_{\rm d}(\nu_0)} \cdot \left(Q_{\rm P}(\nu_0) - Q_{\rm P,CMB}\right)  \nonumber\\
    U_{\rm d}(\nu) &=& \frac{I_{\rm d}(\nu)}{I_{\rm d}(\nu_0)} \cdot \left(U_{\rm P}(\nu_0) - U_{\rm P,CMB}\right),
    \label{QUmodel}
\end{eqnarray}
where $Q_{\rm P,CMB}$ and $U_{\rm P,CMB}$ are CMB Stokes maps from $Planck$.  We use the $Planck$ SMICA CMB maps \citep{Planck2018IV}.
Our {\it Commander} and GNILC models are low resolution versions of dust models in PySM \citep{Thorne17,2021JOSS....6.3783Z}, which are commonly used by the CMB community. 

\subsection{Data simulations}
\label{subsec:data_simulations}
In order to account for the $Planck$ satellite noise including instrumental systematics and CMB signal we compute a set of simulated maps as:
\begin{eqnarray}
\label{eq:qu_ref}
    Q^{\prime}_{\rm sim}(\nu) &= Q_{\rm d}(\nu) + Q_{\rm noise+syst}(\nu) + Q_{\rm CMB}  \nonumber\\
    U^{\prime}_{\rm sim}(\nu) &= U_{\rm d}(\nu) + U_{\rm noise+syst}(\nu) + U_{\rm CMB}, 
\end{eqnarray}
where $[Q_{\rm noise+syst},U_{\rm noise+syst}]$ are the 200 \srolltwo~simulations to which we subtract the \Planck~sky model that contain synchrotron, dust and CMB in order to isolate \emph{Planck} noise/systematics, \cite[see][for more details]{delouis2018}. 
And $[Q_{\rm CMB},U_{\rm CMB}]$ are 200 independent realizations of the CMB computed from the theoretical power spectra of the best-fit $\Lambda$CDM model to the \emph{Planck} data \footnote{We used the data file COM\_PowerSpect\_CMB-base-plikHM-TTTEEE-lowl-lowE-lensing-minimum-theory\_R3.01.txt available on the \emph{Planck} Legacy archive.} \citep{planck2018VI}, with the $r$ parameter equal to 0.  
Thus, Eq.~\ref{eq:qu_ref} yields 200 realizations of the reference model maps. 

\begin{figure}[h!]
    \centering
    \includegraphics[width=0.51\textwidth]{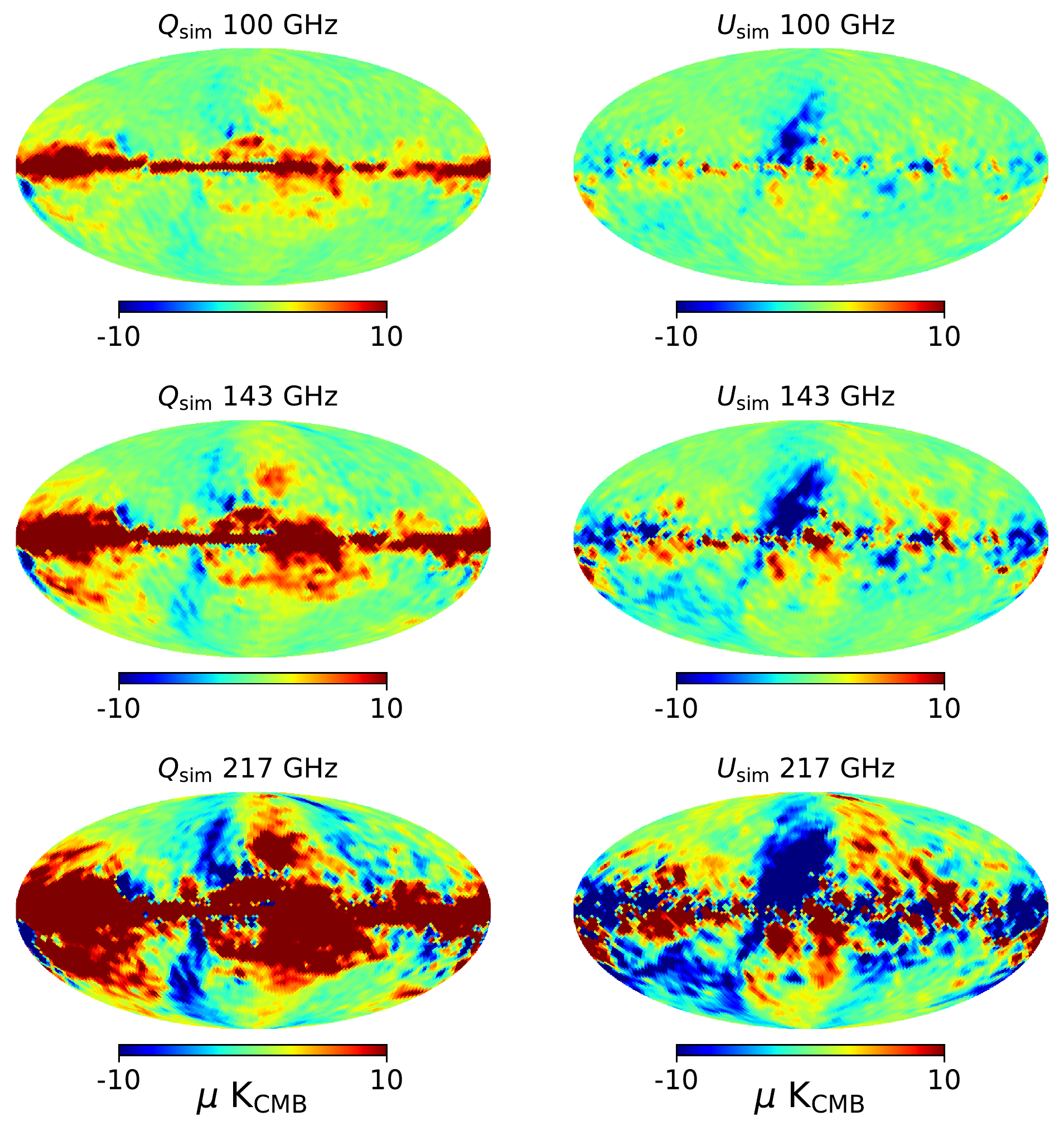}
    \caption{$Q^{\prime}_{\rm sim}$ and $U^{\prime}_{\rm sim}$ reference maps for 100 GHz (top), 143 GHz (middle), and 217 GHz (bottom) as obtained for the {\it Commander} reference model.}
    \label{fig:reference_maps}
\end{figure}

\begin{figure*}[h!]
    \centering
    \includegraphics[width=1\textwidth]{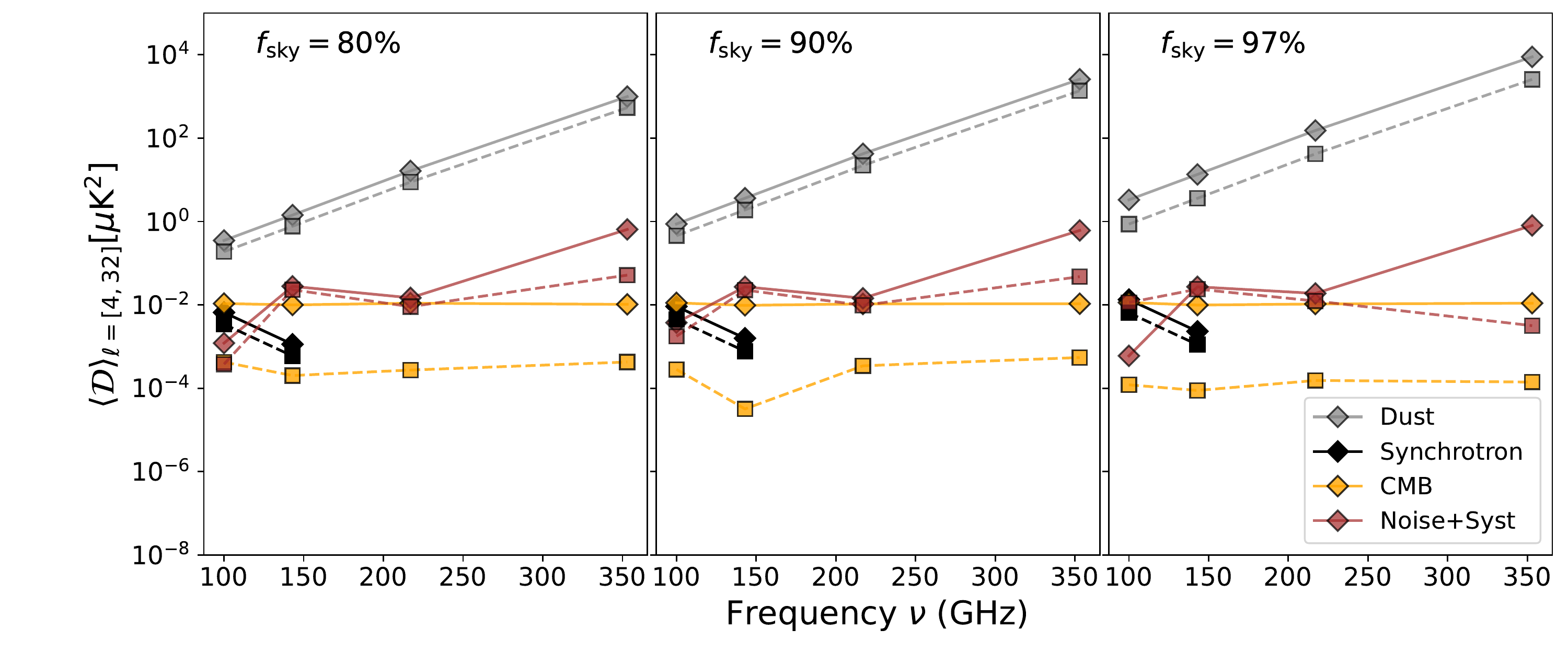}
    \caption{Amplitudes, averaged for 4< $\ell$ <32, of the $\mathcal{D}_\ell$ $EE$ (diamond) and $BB$ (square) power spectra versus frequency. Each plot presents the spectra of the {\it Commander} dust model (gray), our synchrotron estimate at 100 and 143\,GHz (black), the CMB (orange), and the \srolltwo~noise plus systematics (light brown). The three plots from left to right correspond to $f_{\rm sky}=\,$80, 90 and 97\%.} 
    \label{fig:figure1}
\end{figure*}

In Fig.~\ref{fig:reference_maps} we present one realization of the {\it Commander} $Q^{\prime}_{\rm sim}$ and $U^{\prime}_{\rm sim}$ maps at 100, 143, and 217\,GHz. 
In Fig.~\ref{fig:figure1}, the power spectra of the dust model $Commander$, the synchrotron template at the two lowest frequencies, the CMB and a simulation of the \srolltwo\ noise plus systematics are compared. The mean values $\left\langle \mathcal{D}_\ell\right\rangle_{\ell=[4,32]}$ for $EE$ (diamonds) and $BB$ (squares) spectra are plotted  versus frequency  for the three values of $f_{\rm sky}$.
\section{Dust mean SED in polarization}
\label{sec: dust SED}
In this section, we derive the mean SED of dust polarization.
We use $Planck$ polarization maps at 100, 143, 217 and 353\,GHz.
The SED values are normalized to the reference frequency $\nu_0=353$\,GHz.

We follow earlier studies \citep{2016A&A...586A.133P,planck_XI2020} using cross power spectra to determine the dust mean polarized SED for \Planck~data $\gamma_{\rm P}(\nu)$, the reference models $\gamma_{\rm d}(\nu)$ and the simulations $\gamma_{\rm sim}(\nu)$ normalized to $\nu_0$ as:

\begin{eqnarray}
  \begingroup\makeatletter\def\f@size{9}\check@mathfonts
\begin{aligned}
\label{gamma_P}
&  \gamma_{\rm P}^{\rm XX}(\nu) =  \rho_{\nu} \, \left\langle \frac{\mathcal{D}_\ell^{\rm XX}(\nu \times \nu_0)-\mathcal{D}^{\rm XX}_{\ell, \rm CMB}}{\mathcal{D}_\ell^{\rm XX}(\nu_0 \times \nu_0)- \mathcal{D}^{\rm XX}_{\ell, \rm CMB}} \right\rangle_{\ell_{\rm min},\ell_{\rm max}} \\
&   \gamma_{\rm d}^{\rm XX}(\nu) =  \left\langle \frac{\mathcal{D}_{\ell, \rm d}^{\rm XX}(\nu \times \nu_0)}{\mathcal{D}_{\ell,\rm d}^{\rm XX}(\nu_0 \times \nu_0)} \right\rangle_{\ell_{\rm min},\ell_{\rm max}} \\
&   \gamma_{\rm sim}^{\rm XX}(\nu) =  \rho_{\nu} \, \left\langle \frac{\mathcal{D}_{\ell, \rm sim}^{\rm XX}(\nu \times \nu_0)-\mathcal{D}^{\rm XX}_{\ell, \rm CMB}}{\mathcal{D}_{\ell, \rm sim}^{\rm XX}(\nu_0 \times \nu_0)- \mathcal{D}^{\rm XX}_{\ell, \rm CMB}} \right\rangle_{\ell_{\rm min},\ell_{\rm max}}
\end{aligned}
\endgroup
\end{eqnarray}
where 
\begin{eqnarray}
  \begingroup\makeatletter\def\f@size{8.6}\check@mathfonts
\begin{aligned}
&{\mathcal{D}_{\ell}}^{\rm XX} (\nu \times \nu_0) = [{Q^{\prime}_{\rm P}}^{\rm HM1}(\nu), {U^{\prime}_{\rm P}}^{\rm HM1}(\nu)]  \times  [{Q^{\prime}_{\rm P}}^{\rm HM2}(\nu_0), {U^{\prime}_{\rm P}}^{\rm HM2}(\nu_0)] \\
&{\mathcal{D}_{\ell, \rm d}}^{\rm XX} (\nu \times \nu_0) = [{Q_{\rm d}}^{\rm HM1}(\nu), {U_{\rm d}}^{\rm HM1}(\nu)]  \times  [{Q_{\rm d}}^{\rm HM2}(\nu_0), {U_{\rm d}}^{\rm HM2}(\nu_0)] \\
    &{\mathcal{D}_{\ell, \rm sim}}^{\rm XX} (\nu \times \nu_0) = [Q^{\prime~\rm HM1}_{\rm sim}(\nu), U^{\prime~\rm HM1}_{\rm sim}(\nu)]  \times  [Q^{\prime~\rm HM2}_{\rm sim}(\nu_0), U^{\prime~\rm HM2}_{\rm sim}(\nu_0)]
\end{aligned}
\endgroup
\label{dl_planck}
\end{eqnarray}
and $\mathcal{D}_{\ell, \rm CMB}^{\rm XX}$ is the CMB power spectrum for the $\Lambda$CDM fiducial $Planck$ model \citep{planck2018VI} with ${\rm XX}$ $\in$ $[EE,BB]$.
$\mathcal{D}_{\ell}^{\rm XX} (\nu_0 \times \nu_0)$ is computed with $\nu = \nu_0$ in Eq.~\ref{dl_planck}.
The symbol $\times$ indicates the cross-power spectrum operator and $\langle\rangle$ the arithmetic mean over the $\ell-$range from $ \ell_{\rm min}=4$ to $\ell_{\rm max}=32$.
Notice that in the denominator of the first and third equations of Eq.~\ref{gamma_P}, we omit $\rho_{\nu_0}$ because it is equal to 1.
Tab.~\ref{table_mean_sed} lists the SED values $\gamma_{\rm P}(\nu)$ for the \Planck\ data and $\gamma_{\rm d}(\nu)$ for the reference models, for both $EE$ and $BB$ and for the three sky areas.
The uncertainties $\sigma_{\gamma_{\rm P}}(\nu)$ are derived from 
the standard deviation of the 200 $Planck$ simulations. The uncertainty on the polarization efficiencies $\rho_{\nu}$ and the synchrotron spectral index $\beta_{\rm s}$ are propagated through our analysis and their contribution to the total error-bar are added to  $\sigma_{\gamma_{\rm P}}$.

\begin{figure*}[h!]
    \centering
    \includegraphics[width=1\textwidth]{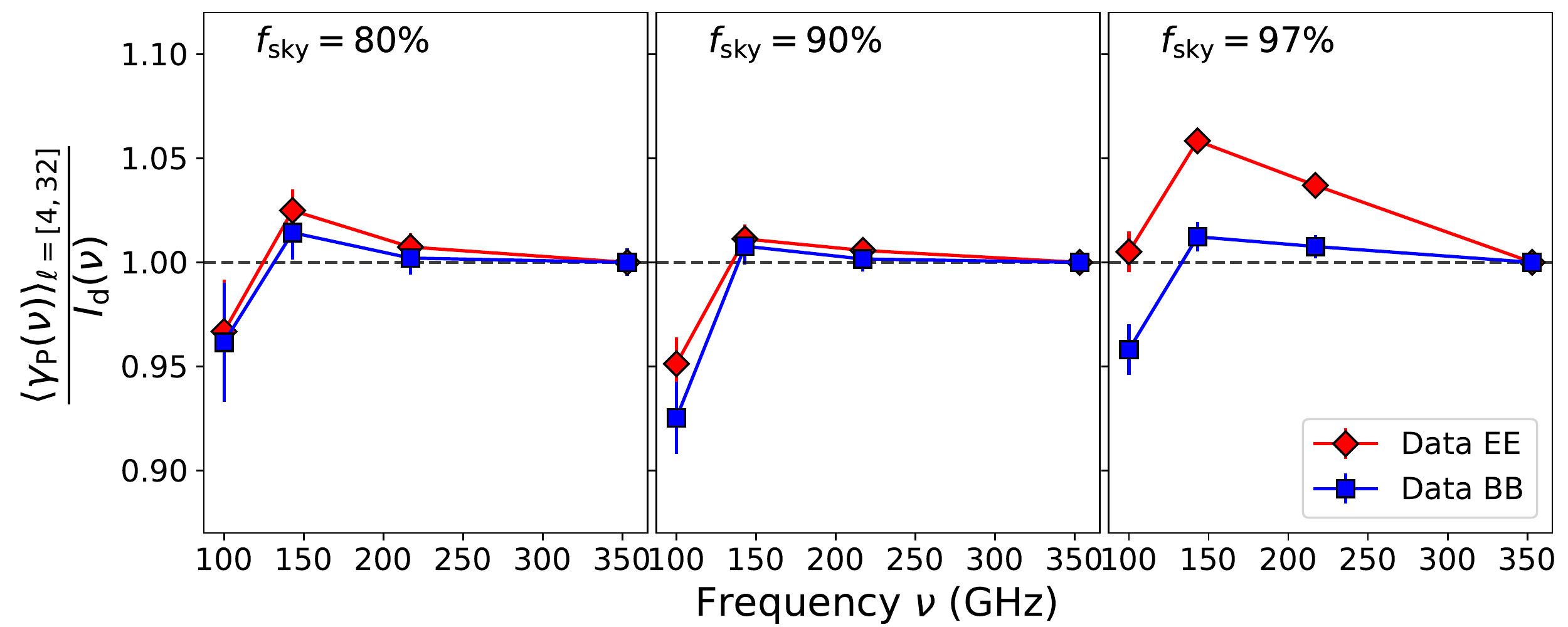}
    \caption{Dust mean SED $\gamma_{\rm P}(\nu)$ normalized by a modified black body function $I_{\rm d}(\nu)$ with fixed $\beta_{\rm d}=1.53$ and $T_{\rm d}=19.6$\,K as given by \cite{planck_XI2020} and accounting for color corrections. $EE$ and $BB$ values are shown in red and blue, respectively. From left to right the results are presented for $f_{\rm sky}$= [80, 90, 97]\%.}
    \label{fig:gammap}
\end{figure*}
\begin{figure*}[h!]
    \centering
    \includegraphics[width=1\textwidth]{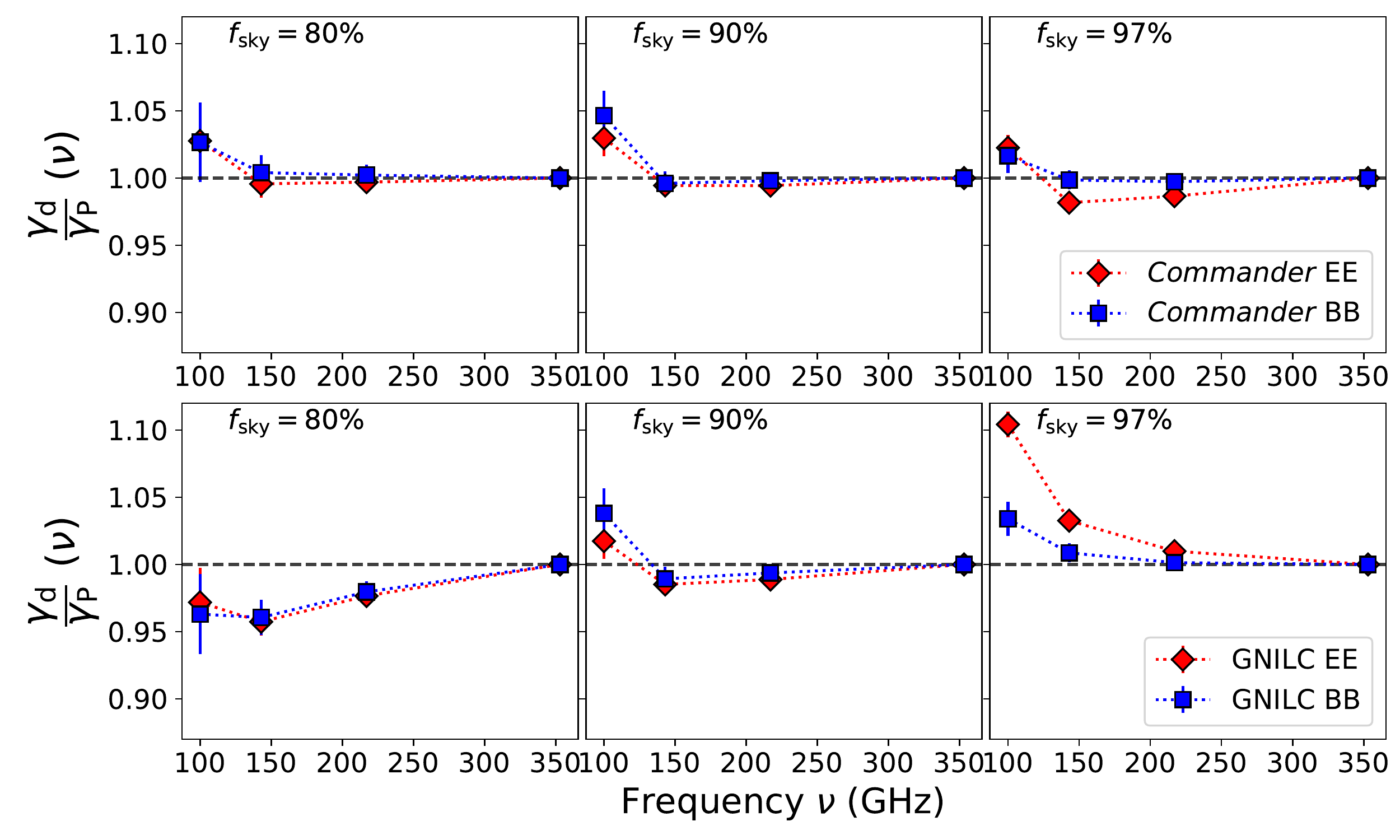}
    \caption{Dust mean SED $\gamma_{\rm d}(\nu)$ computed for the reference models ({\it Commander} (top row) and GNILC (bottom row). $EE$ and $BB$ values are shown in red and blue, respectively. From left to right the results are presented for $f_{\rm sky}$= [80, 90, 97]\%. All values are normalized by the $\gamma_{\rm P}(\nu)$ values obtained from the data and shown in Fig.~\ref{fig:gammap}.}
    \label{fig:gammam}
\end{figure*}

\begin{table*}
\centering
\arrayrulecolor[rgb]{0.753,0.753,0.753}
\label{table_mean_sed}
\begin{tabular}{!{\color{black}\vrule}l!{\color{black}\vrule}l|l|l!{\color{black}\vrule}l|l|l!{\color{black}\vrule}l|l|l!{\color{black}\vrule}} 
\arrayrulecolor{black}\cline{1-10}
\fsky   & \multicolumn{3}{c!{\color{black}\vrule}}{80\%} & \multicolumn{3}{c!{\color{black}\vrule}}{90\%} & \multicolumn{3}{c!{\color{black}\vrule}}{97\%}  \\ 
\hline
$\nu [GHz]$ & 100    & 143    & 217     & 100    & 143    & 217     & 100    & 143    & 217      \\ 
\arrayrulecolor{black}\hline
$\gamma_{\rm P}^{EE}$  & 0.0181 & 0.0400 & 0.1276  & 0.0178 & 0.0392 & 0.1274  & 0.0188 & 0.0411 & 0.1314   \\ 
\arrayrulecolor[rgb]{0.753,0.753,0.753}\hline
$\gamma_{\rm d}^{EE}$\textit{Commander} & 0.0187 & 0.0388 & 0.1291  & 0.0185 & 0.0382 & 0.1286  & 0.0194 & 0.0395 & 0.1315   \\ 
\arrayrulecolor[rgb]{0.753,0.753,0.753}\hline
$\gamma_{\rm d}^{EE}$ GNILC             & 0.0177 & 0.0373 & 0.1265  & 0.0182 & 0.0379 & 0.1279  & 0.0209 & 0.0416 & 0.1347   \\ 
\arrayrulecolor[rgb]{0.753,0.753,0.753}\hline
error                               & 0.0005 & 0.0003 & 0.0005  & 0.0002 & 0.0002 & 0.0003  & 0.0002 & 0.0001 & 0.0002   \\ 
\arrayrulecolor{black}\hline
$\gamma_{\rm P}^{BB}$                   & 0.0180 & 0.0394 & 0.1269  & 0.0173 & 0.0391 & 0.1269  & 0.0180 & 0.0393 & 0.1276   \\ 
\arrayrulecolor[rgb]{0.753,0.753,0.753}\hline
$\gamma_{\rm d}^{BB}$\textit{Commander} & 0.0186 & 0.0387 & 0.1291  & 0.0182 & 0.0382 & 0.1285  & 0.0183 & 0.0384 & 0.1292   \\ 
\arrayrulecolor[rgb]{0.753,0.753,0.753}\hline
$\gamma_{\rm d}^{BB}$GNILC              & 0.0174 & 0.0371 & 0.1262  & 0.0181 & 0.0379 & 0.1279  & 0.0187 & 0.0388 & 0.1297   \\ 
\arrayrulecolor[rgb]{0.753,0.753,0.753}\hline
error                               & 0.0005 & 0.0005 & 0.0008  & 0.0003 & 0.0003 & 0.0004  & 0.0002 & 0.0002 & 0.0003   \\
\arrayrulecolor{black}\hline
\end{tabular}
\caption{Dust mean SED values for the $Planck$ polarization data $\gamma_{\rm P}(\nu)$ and the reference models $\gamma_{\rm d}(\nu)$. The uncertainty is estimated as standard deviation over 200 simulations. For 100 and 143\,GHz the uncertainty accounts for the error associated with the synchrotron template subtraction.}
\end{table*}

The dust SEDs \Planck~$\gamma_{\rm P}$ are presented in Fig.~\ref{fig:gammap}.  The values are normalized to a MBB SED with $\beta_{\rm d}=1.53$ and $T_{\rm d}=19.6$\,K \citep{2014A&A...566A..55P,2015A&A...576A.107P}. Notice that the value at 353 GHz is always equals to unity because it is the reference frequency considered for our analysis, see Eq.~\ref{gamma_P}.
The colour corrections $C^{\rm CC_d}_{\nu}$($\beta_{\rm d}$=1.53 and $T_{\rm d}$=19.6K) are those listed in Tab.~\ref{tab:units}. 
The figure shows that the $\gamma_{\rm P}(\nu)$ values are consistent with the MBB used for normalization within 5\,\%, in agreement with previous results \citep{2015A&A...576A.107P,planck_XI2020}. Some of the differences could be due to systematics that plagues PR3 data. 
Note that subtracting synchrotron emission primarily affects the value of the $EE$ signal to 100 GHz. Though its level is low, around 2 to 3\%.
For the mask at 97\% we observe a significant difference between $\gamma_{\rm P}^{EE}(\nu)$ and $\gamma_{\rm P}^{BB}(\nu)$ that we interpret with variations of the SED within the beam, which do not average in the same way for $EE$ and $BB$. These variations depend also on the mask and could be more important when the analysis includes a significant part of the Galactic plane.

Fig.~\ref{fig:gammam} shows the ratio between $\gamma_{\rm d}$ for the \textit{Commander} (top) and GNILC (bottom) models and $\gamma_{\rm P}$.
For {\it Commander} we observe a very close match between $\gamma_{\rm P}(\nu)$ and $\gamma_{\rm d}(\nu)$ at 143 and 217 GHz for $f_{\rm sky}$=80 and 90 \% and both $EE$ and $BB$ signals. For GNILC the match is not as good, in particular we observe a significant difference for \fsky=80\% at 100 and 143 GHz.
For $f_{\rm sky}$=97\% we observe a difference between the $EE$ and $BB$ values at 100 GHz.
Apart for this latter we observe that $\gamma_{\rm d}(\nu)$ values are consistent within 5\% with the $\gamma_{\rm P}(\nu)$ values.
All the differences between $\gamma_{\rm P}$ $\gamma_{\rm d}$ may be due to averaging effects along the line of sight.


This analysis shows that assuming a reference model based on total intensity data could not completely reproduce the SED observed in polarization data, thus biasing any CMB polarization \emodes~and~\bmodes~signals if used as reference for component separation methods. In order to ensure an unbiased detection of the CMB polarization, at the precision required from future CMB experiments, we need to address these very small spatial SED variations of the dust polarization emission.

\section{Spatial variations of the polarization SED}
\label{sec:spatial_variations}
In this section, we characterize the spatial variations of the dust SED in polarization and quantify the degree of correlation with variations of the dust SED in total intensity. 

\subsection{Residual maps}
\label{sec:residual_maps}
\begin{figure}[h!]
  \centering
  \includegraphics[width=0.51\textwidth]{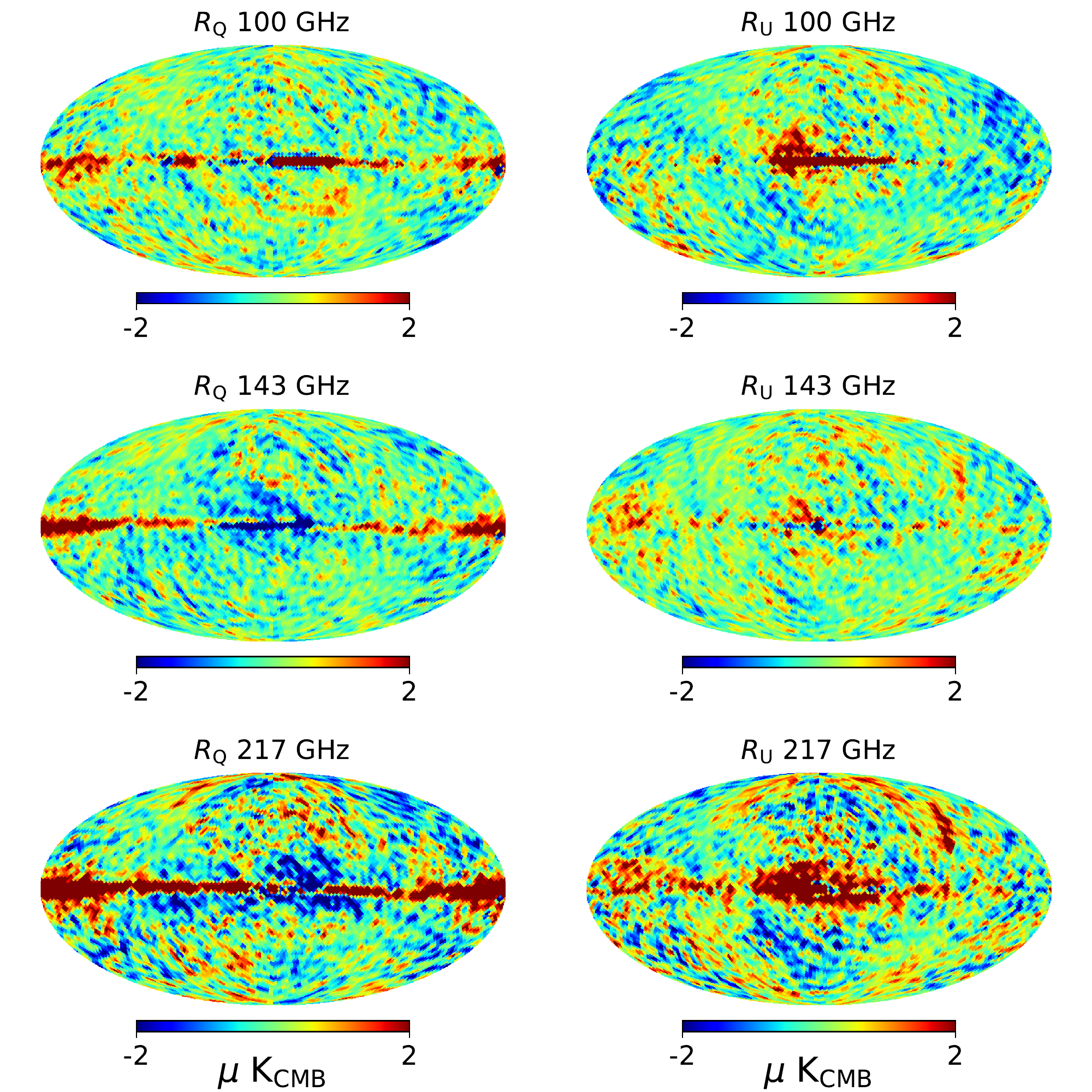}
\caption{From top to bottom residual maps $R_Q$ (left) and $R_U$ (right) at 100, 143 and 217\,GHz.}
\label{fig:residuals_qu} 
\end{figure}

\begin{figure}[h!]
  \centering
  \includegraphics[width=0.51\textwidth]{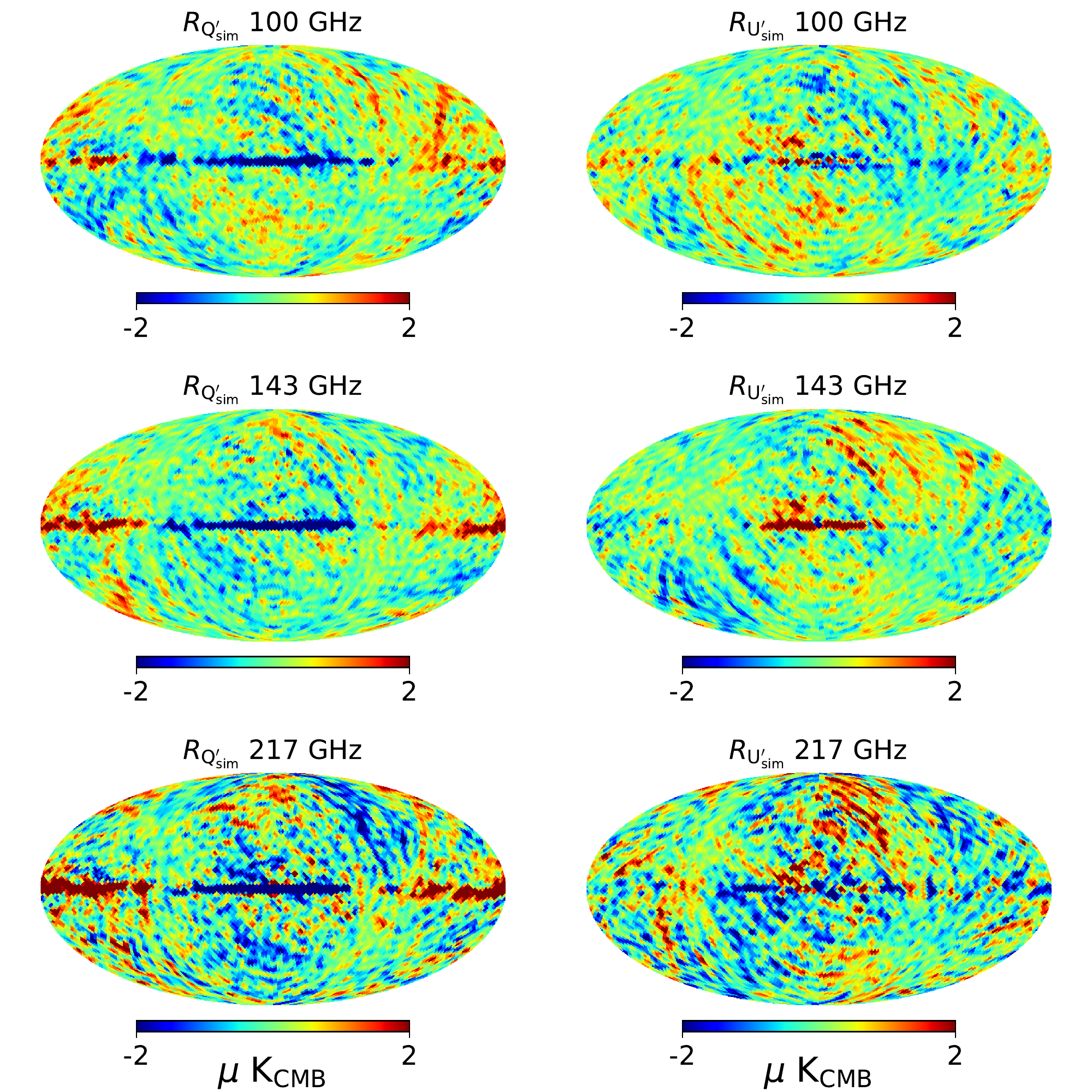}
  \caption{From top to bottom residual maps for one realization of $R_{Q^{\prime}_{\rm sim}}$ (left) and $R_{U^{\prime}_{\rm sim}}$ (right) obtained for the {\it Commander} reference model, at 100, 143, 217 GHz respectively.}
\label{fig:residuals_model_qu}
\end{figure}

\begin{figure}[h!]
  \centering
  \includegraphics[width=0.5\textwidth]{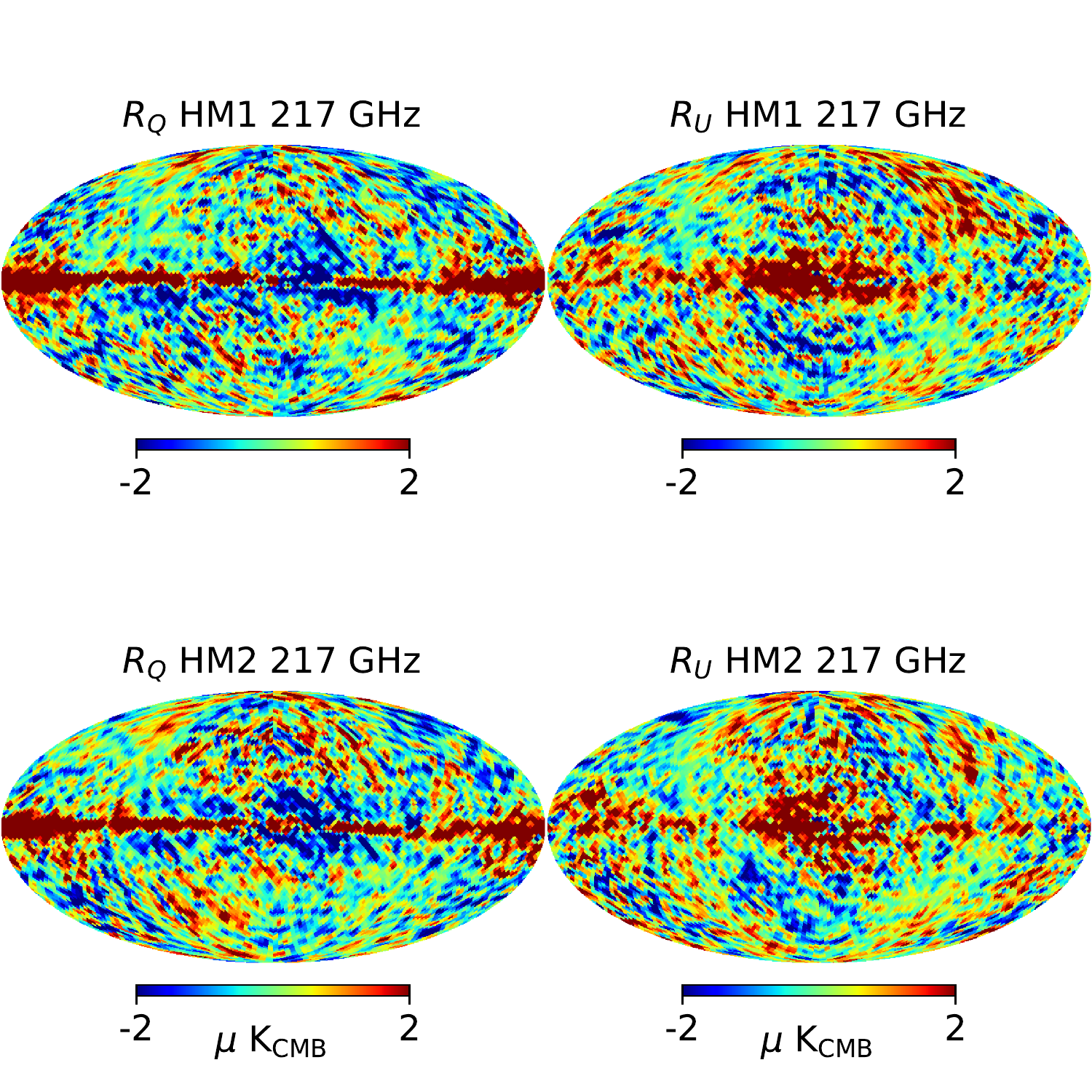}
\caption{From left to right, residual maps $R_{\rm Q}$ (left) and $R_{\rm U}$ (right) at 217 GHz for two data sets HM1 (top), HM2 (bottom), respectively.}
\label{hm1_2_residual_maps} 
\end{figure}

To quantify the spatial variations of the dust SED, we compute differences between the $Planck$ frequency maps at 100, 143 and 217\,GHz and the $353\,$GHz scaled by the mean dust SED $\gamma_{\rm P}(\nu)$ from Sect.~\ref{sec: dust SED}, that we approximate to the value of $\gamma_{\rm P}^{EE}(\nu)$ because the dust polarization is dominated by \emodes~.  Fig.~\ref{fig:gammap} shows that $\gamma_{\rm P}^{EE}(\nu) \simeq \gamma_{\rm P}^{BB}(\nu)$ for $f_{\rm sky}$=80\%, 90\%. A difference between the two coefficients $\gamma_{\rm P}(\nu)$ is instead detected for $f_{\rm sky}$=97\%. This difference indicates a correlation between dust emission properties and the structure of the magnetized interstellar medium. However, we have checked that our approximation does not significantly impact the results also for this $f_{\rm sky}$.
The three maps $R_Q(\nu)$ and $R_U(\nu)$ computed as:
\begin{eqnarray}
    \label{eq:resid_maps}
   R_{\rm Q}(\nu) &=  Q^{\prime}_{\rm P} (\nu) - \gamma_{\rm P}(\nu) \cdot Q_{\rm P} (\nu_0) \\ \nonumber
   R_{\rm U}(\nu) &=  U^{\prime}_{\rm P} (\nu) - \gamma_{\rm P}(\nu) \cdot U_{\rm P} (\nu_0)
\end{eqnarray} 
with $\nu$= 100, 143 and 217 GHz, and $\nu_0=353$ GHz to which hereafter we refer as residual maps, are shown in Fig.~\ref{fig:residuals_qu}. 
For comparison, we also compute difference maps for the reference models (see Eq.~\ref{eq:res_maps_model}) and simulations maps replacing $\gamma_{\rm P}(\nu)$ by $\gamma_{\rm d}(\nu)$ in Eq.~\ref{eq:resid_maps}. 
\begin{eqnarray}
\begin{aligned} 
&R_{\rm Q_{\rm d}}(\nu) = Q_{\rm d} (\nu) - \gamma_{\rm d}(\nu) \cdot Q_{\rm d} (\nu_0) \\ 
&R_{\rm U_{\rm d}}(\nu) = U_{\rm d} (\nu) - \gamma_{\rm d}(\nu) \cdot U_{\rm d} (\nu_0) \\ 
&R_{\rm Q^{\prime}_{sim}}(\nu) = Q^{\prime}_{\rm sim} (\nu) - \gamma_{\rm d}(\nu) \cdot Q^{\prime}_{\rm sim} (\nu_0) \\
&R_{\rm U^{\prime}_{sim}}(\nu) = U^{\prime}_{\rm sim} (\nu) - \gamma_{\rm d}(\nu) \cdot U^{\prime}_{\rm sim} (\nu_0) 
\label{eq:res_maps_model}
\end{aligned} 
\end{eqnarray}
Notice that, for illustration purpose, all the residual maps shown in Fig.~\ref{fig:residuals_qu}\,,\ref{fig:residuals_model_qu}\,,\ref{hm1_2_residual_maps} consider a $\gamma_{\rm P}(\nu)$ and $\gamma_{\rm d}(\nu)$ in Eq.~\ref{eq:resid_maps} computed for $f_{\rm sky}=$90\%.

One set of simulation maps [R$_{\rm Q^{\prime}_{\rm sim}}$($\nu$), R$_{\rm U^{\prime}_{\rm sim}}$($\nu$)] for the {\it Commander} model  is displayed in 
Fig.~\ref{fig:residuals_model_qu}. The comparison between  Figs.~\ref{fig:residuals_qu} and \ref{fig:residuals_model_qu}
is hampered by $Planck$ data noise but one can notice some common features and some differences. 
In Fig.~\ref{fig:residuals_qu}, we also note differences between frequencies among the $R_{\rm Q}$\,and $R_{\rm U}$  residual maps, in contrast to what is observed for $R_{\rm Q^{\prime}_{\rm sim}}$\,and $R_{\rm U^{\prime}_{\rm sim}}$ in Fig.~\ref{fig:residuals_model_qu}. 
To illustrate that some of the structures are above data noise and uncorrected systematics, in Fig.~\ref{hm1_2_residual_maps} we present the independent $R_{\rm Q}$\,and $R_{\rm U}$ residual half-mission HM1\,and HM2 maps estimated at 217\,GHz.

This qualitative examination of the maps suggests that we observe SED variations in the residual $R_{\rm Q}(\nu)$, $R_{\rm U}(\nu)$ maps, which cannot be simply attributed to SED variations in total intensity, as assumed in the reference models. 
\begin{figure*}[h!]
  \centering
  \includegraphics[width=1\textwidth]{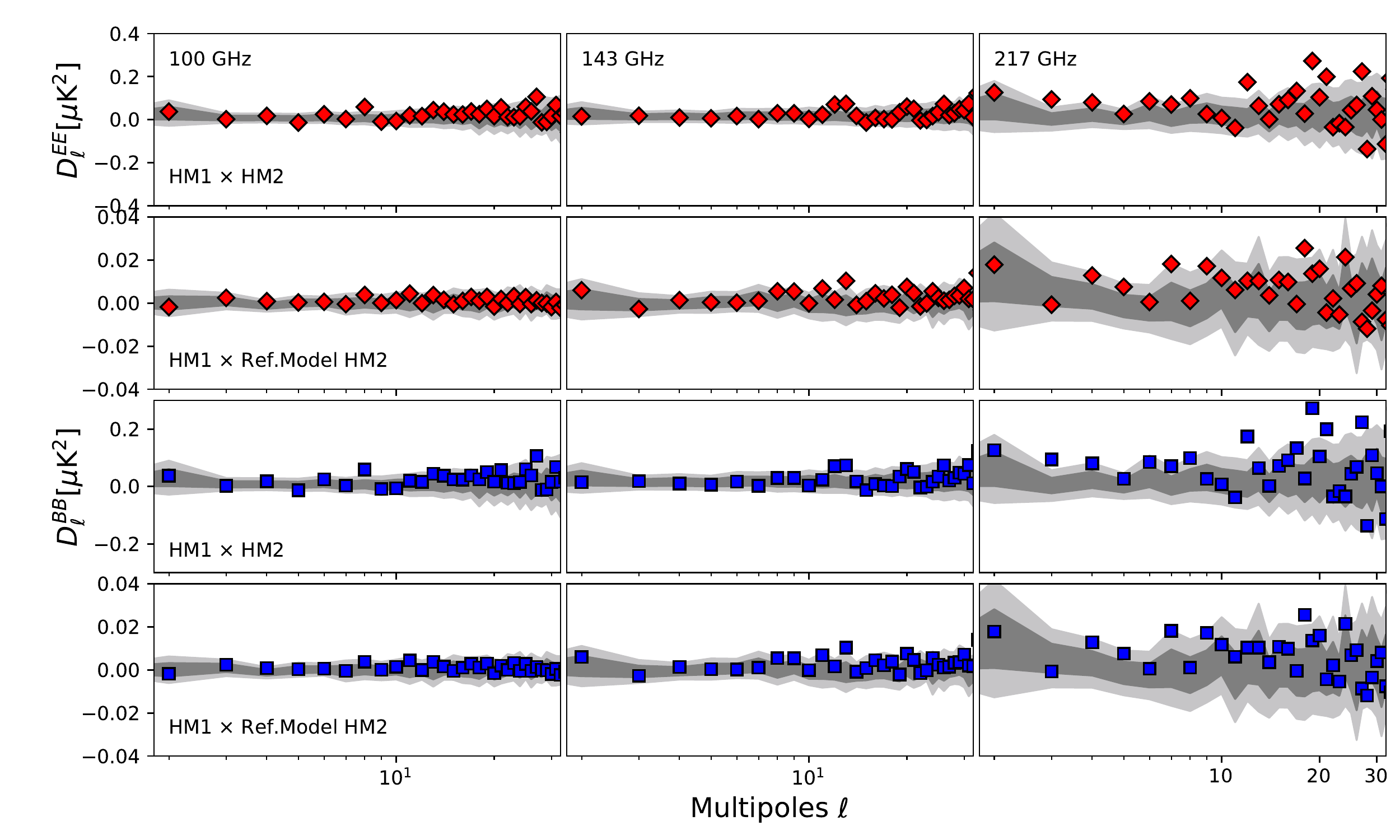}
\caption{Cross power spectra correlation  $\mathcal{D}_{\ell}^{\rm XX}(\nu)$ and $\mathcal{D}_{\ell,\rm d}^{\rm XX}(\nu)$ indicated as HM1 $\times$ HM2 and HM1 $\times$ Ref.model HM2, respectively, in the labels. Results for $f_{\rm sky}$=90\% and {\it Commander} reference model are shown. From top to bottom the first two rows show the $\mathcal{D}_{\ell}^{EE}$ and the last two the $\mathcal{D}_{\ell}^{BB}$ cross power spectra, respectively. From left to right the results at at 100, 143, and 217 GHz are shown. The dark and light gray shades represent the 1$\sigma$ and 2$\sigma$ standard deviation computed from the data simulations.}
\label{fig:spectra_hm1hm2_model} 
\end{figure*}

\subsection{Power spectra analysis}
\label{subsec:residuals_spectra}

To characterize SED variations, we compute power spectra  of residual maps and cross-power spectra with the reference models. 
We compute two cross-power spectra between (i) independent residual maps [$R_{\rm Q}^{\rm HM1,HM2} (\nu)$, $R_{\rm U}^{\rm HM1,HM2} (\nu) $], and (ii) between [$R_{\rm Q}^{\rm HM1} (\nu)$, $R_{\rm U}^{\rm HM1}(\nu)$] and the reference model residuals [$R_{\rm Q_d}^{\rm HM2} (\nu)$, $R_{\rm U_d}^{\rm HM2} (\nu)$]. The 200 simulations are used to assess the impact of data noise plus uncorrected systematics and of the CMB on our analysis.
\begin{eqnarray}
\begin{aligned}
& \mathcal{D}_{\ell, \rm res}^{\rm XX} (\nu) = [R_{\rm Q}^{\rm HM1}(\nu), R_{\rm U}^{\rm HM1}(\nu)]  \times  [R_{\rm Q}^{\rm HM2}(\nu), R_{\rm U}^{\rm HM2}(\nu)] \\ 
& \mathcal{D}_{\ell, \rm d, \rm res}^{\rm XX} (\nu) = [R_{\rm Q}^{\rm HM1}(\nu), R_{\rm U}^{\rm HM1}(\nu)]  \times  [R_{\rm Q_d}^{\rm HM2}(\nu), R_{\rm U_d}^{\rm HM2}(\nu)]\\ 
& \mathcal{D}_{\ell, \rm sim, \rm res}^{\rm XX} (\nu) = [R_{\rm Q^{\prime}_{\rm sim}}^{\rm HM1}(\nu), R_{\rm U^{\prime}_{\rm sim}}^{\rm HM1}(\nu)]  \times  [R_{\rm Q^{\prime}_{\rm sim}}^{\rm HM2}(\nu), R_{\rm U^{\prime}_{\rm sim}}^{\rm HM2}(\nu)]
 \end{aligned}
\end{eqnarray}
where ${\rm XX}$ $\in$ $[EE, BB]$.
Fig.~\ref{fig:spectra_hm1hm2_model} shows 
both sets of cross power spectra for $EE$ (red) and $BB$ (blue) at 100, 143 and 217\,GHz for $f_{\rm sky} = 90\,\%$. 
The data points over the $\ell$-range 4 to 32 are compared with the 1$\sigma$ and 2$\sigma$ dispersion (dark and light gray shades) obtained with the 200 simulations including noise and CMB anisotropies (see Sect.~\ref{subsec:data_simulations}). 
In all plots, individual data points have a low signal-to-noise ratio and it is necessary to average them to quantify mean amplitudes and their frequency dependence. 
\begin{figure*}[h!]
  \centering
  \includegraphics[width=1\textwidth]{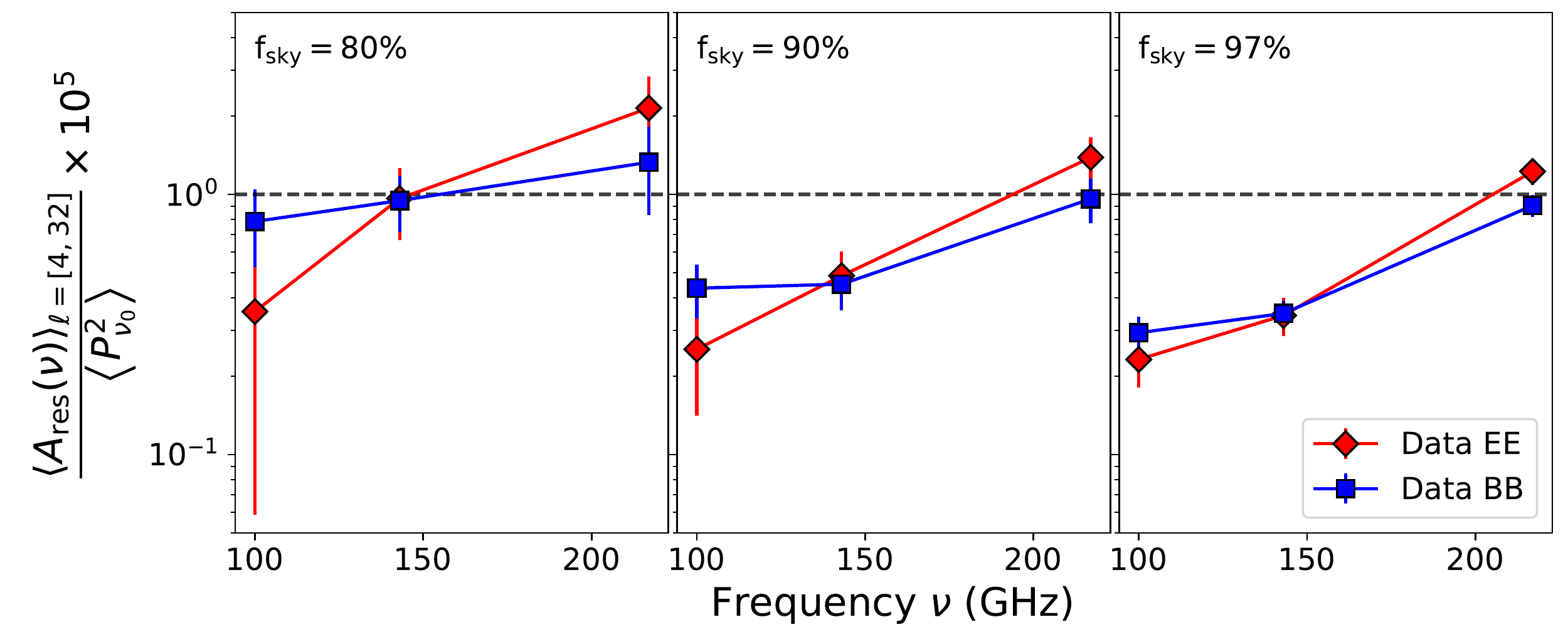}
\caption{ Amplitudes $A_{\rm res}(\nu)$ for $EE$ (red)\,and $BB$ (blue), respectively. These values are averages over the $\ell$-range 4 to 32 and normalized by the mean of $P_{\nu_0}^2$. From left to right results for $f_{\rm sky}$=[80, 90, 97]\% are shown.}
\label{fig:sed_variationp} 
\end{figure*}
\begin{figure*}[h!]
  \centering
  \includegraphics[width=1\textwidth]{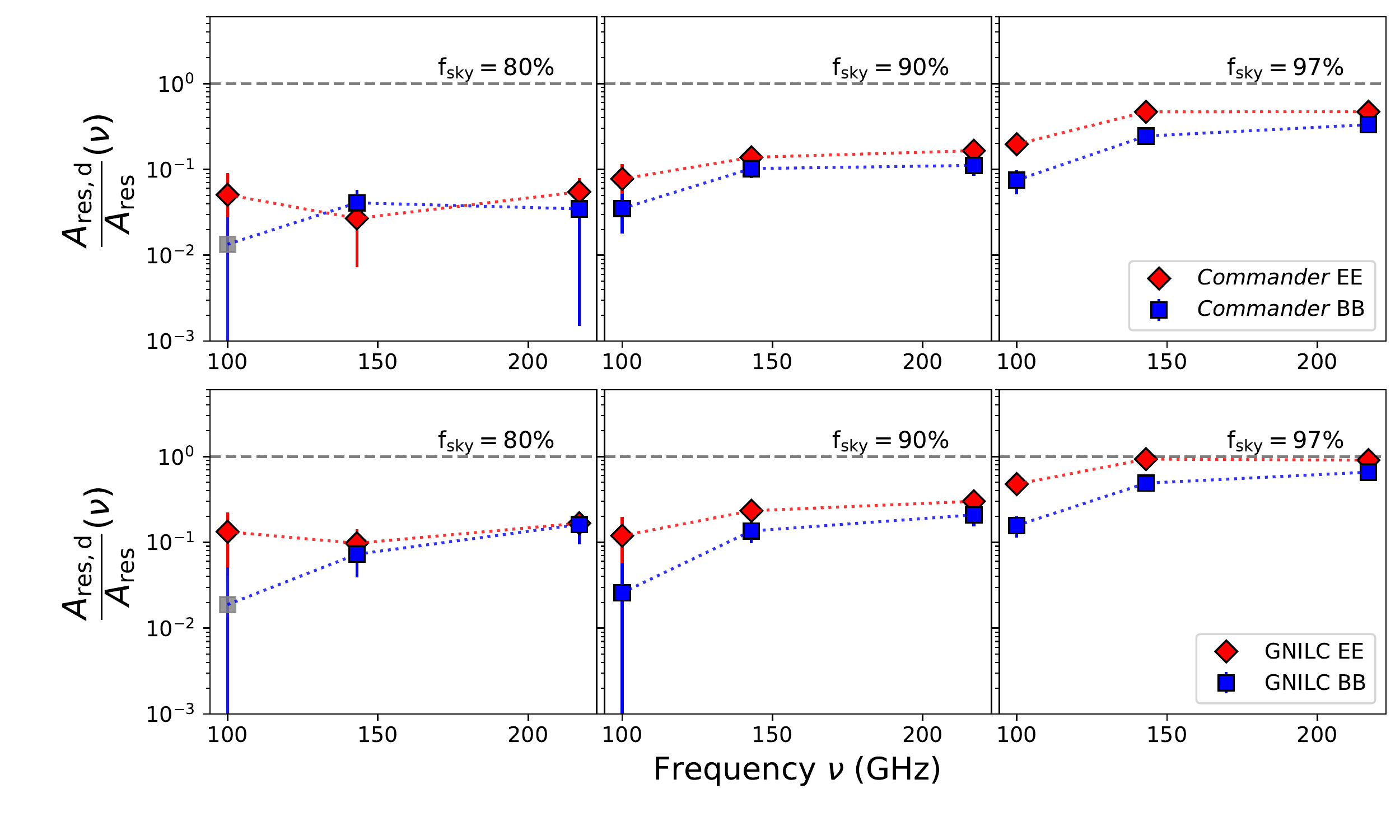}
\caption{Residual amplitudes $A_{\rm res, d}(\nu)$ obtained for the reference models $Commander$ (top) and GNILC (bottom). Red and blue colors represent the $EE$\,and $BB$ results, respectively. These values are estimated by averaging between $\ell$-range 4 to 32 and normalizing to the amplitudes of the residuals obtained from the \Planck~data $A_{\rm res}(\nu)$. See Eq.~\ref{eq:ares} for details. From left to right results for $f_{\rm sky}$=[80, 90, 97]\% are shown. In gray color it is represented the absolute value of $BB$ negative results found at 100 GHz for $f_{\rm sky}$=80\%.}
\label{fig:sed_variationm} 
\end{figure*}
\begin{figure*}[h!]
  \centering
 \includegraphics[width=1\textwidth]{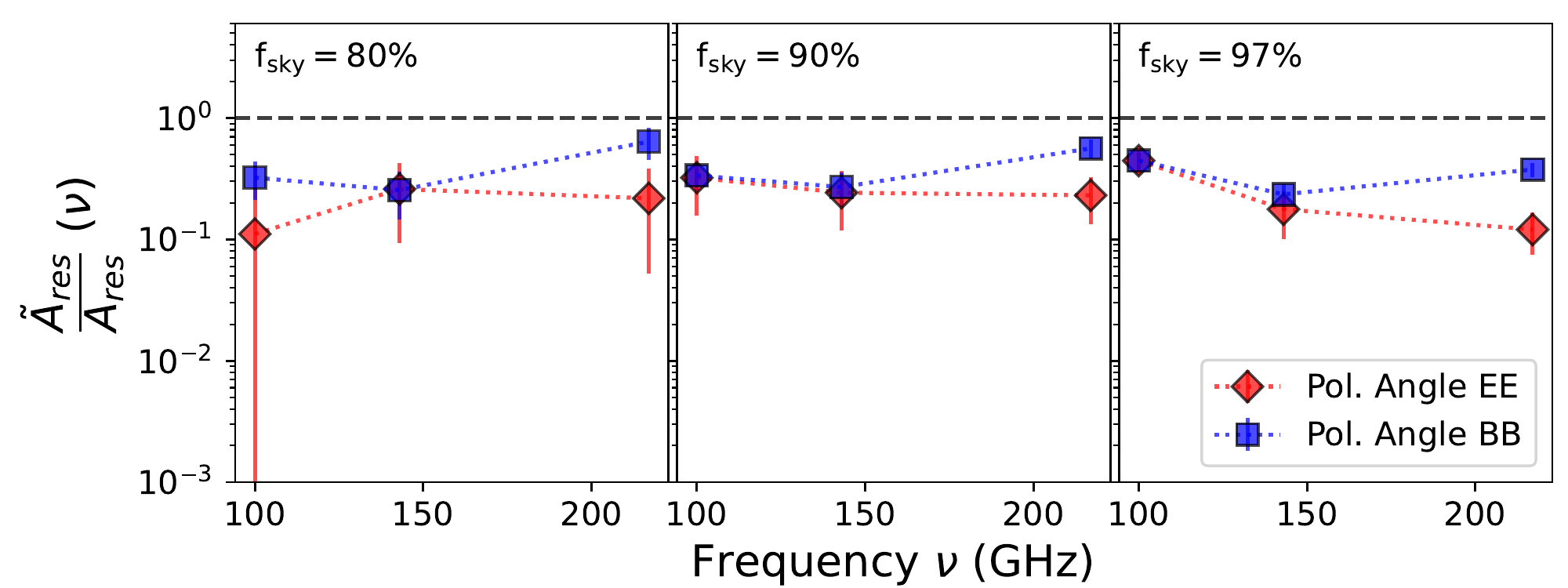}
 \caption{Amplitudes of the residuals $\Tilde{A}_{\rm res}(\nu)$ obtained by isolating the effect of the polarization angle variation are shown for $EE$ (red) and $BB$ (blue) (see Eq.~\ref{eq:ares_angle}). These values are normalized to the residual amplitudes obtained from the \Planck~data analysis ${A}_{\rm res}(\nu)$. From left to right values for $f_{\rm sky}$=80, 90, and 97 \% are shown.}
 \label{fig:sed_variation_angle}
 \end{figure*}
The averaged cross power spectra values are computed as:
\begin{eqnarray}
 \begingroup\makeatletter\def\f@size{9}\check@mathfonts
\begin{aligned}
\label{eq:ares}
&A_{\rm res}(\nu) = \left\langle \mathcal{D}_{\ell,\rm res}^{\rm XX}(\nu)- (1-\gamma_{\rm P}(\nu))^2\cdot \mathcal{D}_{\ell,\rm CMB}^{\rm XX} \right\rangle_{\rm \ell_{min},\ell_{max}} \\ 
&A_{\rm res, d}(\nu) = \left\langle \mathcal{D}_{\ell, \rm d, res}^{\rm XX}(\nu) \right\rangle_{\rm \ell_{min},\ell_{max}} \\
&A_{\rm res,sim}(\nu) = \left\langle \mathcal{D}_{\ell,\rm res,sim}^{\rm XX}(\nu)- (1-\gamma_{\rm P}(\nu))^2\cdot \mathcal{D}_{\ell, \rm CMB}^{\rm XX}  \right\rangle_{\rm \ell_{min},\ell_{max}} 
\end{aligned}
\endgroup
\end{eqnarray}
with ${\rm XX}$ $\in$ $[EE,BB]$. Notice that the CMB has been partially subtracted in Eqs.~\ref{eq:resid_maps}, \ref{eq:res_maps_model}. The remaining contribution is accounted for by the term $(1-\gamma_{\rm P}(\nu))^2\cdot \mathcal{D}_{\ell, \rm CMB}^{\rm XX}$. 
In Fig.~\ref{fig:sed_variationp} we plot the amplitudes $A_{\rm res} (\nu)$ normalized by the mean polarized intensity at reference frequency over each sky area defined as: 
\begin{eqnarray}
\label{p0}
P_{\nu_0} =\sqrt{|Q_{HM1}\cdot Q_{HM2}+U_{HM1}\cdot U_{HM2}|}
\end{eqnarray}
where HM1 and HM2 refer to the two half-mission maps. We checked that the bias on this estimator of the polarized intensity is negligible at 353\,GHz for \Nside=32. 
This normalization allows us to compare the results for the three different masks. Absolute values of the residuals may be obtained by scaling the figure data points by $\langle P_{\nu_0}^2\rangle$ = [2453.6, 5950.5, 13814.3] K$_{\rm CMB}$ for \fsky = 80\%, 90\%, and 97\% respectively. We will discuss the frequency dependence later, here we focus on the comparison with the models.
Fig.~\ref{fig:sed_variationm} shows the dust models residuals $A_{\rm res, d} (\nu)$ normalized by $A_{\rm res} (\nu)$.
The mean amplitudes $A_{\rm res, sim} (\nu)$ are also computed for each of the 200 simulations and their dispersion provides the error bars. 
For the 80 and 90\% masks, the $A_{\rm res} (\nu)$ are a factor of about 10 times larger than $A_{\rm res,d} (\nu)$. For these, uncertainties associated with the chance correlation between the CMB and the dust residuals may thus be somewhat underestimated. 
For the 97\% mask, this difference is  smaller. At 143 and 217\,GHz, the amplitudes even coincide for the GNILC model. These results show that the reference model only accounts for a minor fraction of the total polarization SED variations in the 80 and 90\% masks but a much more significant one for the brighter dust emission in the Galactic plane.
Furthermore Fig.~\ref{fig:sed_variationp} highlights the significant variation of $BB$ residuals at 100 GHz toward higher galactic latitudes. Although both $EE$ and $BB$ variation are very small, in the case of $f_{\rm sky}$=80\% $A_{\rm res}^{BB}$ represents 0.4\% of the total amplitude w.r.t 0.1\% detected for $A_{\rm res}^{EE}$. This feature does not match the residuals extrapolated from the models considered, see Fig.~\ref{fig:sed_variationm} for comparison.

To give a more quantitative estimate we translate the amplitude $A_{\rm res}(\nu)$ in terms of an effective dispersion of the dust spectral index $\sigma_{\beta}$, using the following formula based on a first order expansion of the MBB emission law in \cite{mangilli2021}.
\begin{eqnarray}
\begin{aligned}
\label{sigmabeta}
&\sigma_{\beta} (\nu) = \left(A_{\rm res}^{EE}(\nu)+A_{\rm res}^{BB}(\nu)\right)^{0.5} \cdot \left(\gamma_{\rm P}(\nu)\cdot P_{\nu_0}\cdot\left|\ln{\left(\frac{\nu}{\nu_0}\right)}\right|\right)^{-1} 
\end{aligned}
\end{eqnarray}
The values of $\sigma_{\beta}$ are listed in Table~\ref{tab:sigbeta_deltapsi} for the three sky areas and frequencies. 
These numbers of about 0.1 must be considered as an estimate of the variance of residuals in terms of pure variation of $\beta$.
These results are consistent with previous studies at low latitude \citep{GNILC}. 
\begin{table*}[h!]
\centering
\arrayrulecolor[rgb]{0.753,0.753,0.753}
\scalebox{0.95}{\begin{tabular}{!{\color{black}\vrule}l!{\color{black}\vrule}l|l|l!{\color{black}\vrule}l|l|l!{\color{black}\vrule}l|l|l!{\color{black}\vrule}} 
\arrayrulecolor{black}\cline{1-10}
\fsky   & \multicolumn{3}{c!{\color{black}\vrule}}{80\%} & \multicolumn{3}{c!{\color{black}\vrule}}{90\%} & \multicolumn{3}{c!{\color{black}\vrule}}{97\%}  \\ 
\hline
$\nu [GHz]$ & 100 & 143 & 217 & 100 & 143 & 217 & 100 & 143 & 217 \\
\arrayrulecolor{black}\hline
$\sigma{_\beta}$ & 0.15$\pm$0.03 & 0.12 $\pm$ 0.01 & 0.09$\pm$0.02 & 0.12$\pm$0.02 & 0.09$\pm$0.01 & 0.08$\pm$0.01& 0.10$\pm$0.01 & 0.07 $\pm$ 0.01 & 0.07 $\pm$ 0.01   \\
\arrayrulecolor{black}\hline
$\delta{_\psi}$ & 2.71$\pm$0.46 & 1.60 $\pm$ 0.25 & 0.81$\pm$0.19 & 2.43$\pm$0.29 & 1.13$\pm$0.15 & 0.66$\pm$0.11& 2.33$\pm$0.20 & 0.83 $\pm$ 0.09 & 0.48 $\pm$ 0.07 \\
\arrayrulecolor{black}\hline
\end{tabular}}
\caption{Effective dispersion of the dust spectral index $\sigma_{\beta}$ and effective angle variation $\delta_{\psi}$. Notice that a systematic uncertainty on the polarization angle of 1$^{\circ}$ must be considered as upper limit on the $Planck$ polarization absolute accuracy \citep{rosset}.}
\label{tab:sigbeta_deltapsi}
\end{table*}


\section{Residuals from polarization angles}
\label{sec:dust_polarization_angle}
In this section, we quantify variations of the dust polarization angles as a function of frequency.
To do this we introduce the maps
\begin{eqnarray}
\label{qu_angle}
\Tilde{Q}(\nu) &= \gamma_{\rm P}(\nu)\cdot P({\nu_0})\times \cos{(2\psi(\nu))} \\ \nonumber
\Tilde{U}(\nu) &= \gamma_{\rm P}(\nu)\cdot P ({\nu_0})\times \sin{(2\psi(\nu))}
\end{eqnarray}
where $P({\nu_0})$ is the polarized intensity as estimated in Eq.\ref{p0}, 
and  $\psi(\nu)=\frac{1}{2}\arctan{\left(\frac{U}{Q}\right)}$ is the polarization angle. 

To assess the contribution of  variations of $\psi$ to the total $A_{\rm res}(\nu)$ values, 
we compute residual maps also for $\Tilde{Q}(\nu)$\, and $\Tilde{U}(\nu)$ as:
\begin{eqnarray}
\label{qu_angle}
\Tilde{R}_{\rm Q}(\nu) &= \Tilde{Q}(\nu)-\gamma_{\rm P}(\nu)\cdot \Tilde{Q}(\nu_0) \\ \nonumber
\Tilde{R}_{\rm U}(\nu) &= \Tilde{U}(\nu)-\gamma_{\rm P}(\nu)\cdot \Tilde{U}(\nu_0)
\end{eqnarray}

The power spectra analysis is performed with these residual maps as described in the previous section:
\begin{eqnarray}
\begin{aligned}
\label{dl_angle}
&\Tilde{\mathcal{D}}_{\ell, \rm res}^{\rm XX} (\nu) = [\Tilde{R}_{\rm Q}^{\rm HM1}(\nu), \Tilde{R}_{\rm U}^{\rm HM1} (\nu)] \times  [\Tilde{R}_{\rm Q}^{\rm HM2}(\nu), \Tilde{R}_{\rm U}^{\rm HM2}(\nu)] \\
&\Tilde{\mathcal{D}}_{\ell,\rm sim, \rm res}^{\rm XX} (\nu) = [\Tilde{R}_{\rm Q^{\prime}_{\rm sim}}^{\rm HM1}(\nu), \Tilde{R}_{\rm U^{\prime}_{\rm sim}}^{\rm HM1} (\nu)] \times  [\Tilde{R}_{\rm Q^{\prime}_{\rm sim}}^{\rm HM2}(\nu), \Tilde{R}_{\rm U^{\prime}_{\rm sim}}^{\rm HM2}(\nu)] 
\end{aligned}
\end{eqnarray}

And then the averaged amplitudes are defined as:
\begin{eqnarray}
\begin{aligned}
\label{eq:ares_angle}
\Tilde{A}_{\rm res}(\nu) = \left\langle \Tilde{\mathcal{D}}_{\ell, \rm res}^{\rm XX}(\nu) - 
\left\langle \Tilde{\mathcal{D}}_{\ell,\rm sim, res}^{\rm XX}(\nu)\right\rangle_{\rm sim}\right\rangle_{\rm \ell_{min},\ell_{max}}, 
\end{aligned}
\end{eqnarray}
where the second term is used for the CMB and residual noise and systematics debiasing.
As in Sect.~\ref{subsec:residuals_spectra} for $A_{\rm res}$, the uncertainties on 
$\Tilde{A}_{\rm res}$ are derived from the dispersion of the values obtained for the simulations. 

Figure~\ref{fig:sed_variation_angle} compares $\Tilde{A}_{\rm res} (\nu)$ values to $A_{\rm res} (\nu)$.
We find that variations of the polarization angle contribute significantly to the total polarization  residuals. 
The differences observed with $f_{\rm sky}$ may result from the integration along the line of sight or from the limit of our debiasing method, or both. Indeed, for decreasing Galactic latitudes, the line of sight crosses an increasing number of coherent turbulent cells, and the impact of the magnetic field structure on observed polarization angles may average out. 


To express the amplitude $\Tilde{A}_{\rm res}(\nu)$ in terms of an effective angle variation $\delta_{\psi}$, we assume that the residuals results from a systematic angle change over the full sky area. This calculation is just indicative because we do not think that this is a valid assumption. Using equations detailed by \citet{Abitbol2016}, we obtain:   
\begin{eqnarray}
\begin{aligned}
\label{eq:deltapsi}
&\sin{\delta_{\psi}} (\nu) = 0.5\, \left(\Tilde{A}_{\rm res}^{EE}(\nu)+\Tilde{A}_{\rm res}^{BB}(\nu)\right)^{0.5} \, \gamma_{\rm P}^{-1}(\nu) (\langle P_{\nu_0}^2\rangle)^{-0.5}
\end{aligned}
\end{eqnarray}
The values of $\delta_{\psi}$ are listed in Table~\ref{tab:sigbeta_deltapsi} for three sky areas and frequencies. These values are small but larger than the $1^\circ$ uncertainty on the ground calibration of the $Planck$ absolute polarization angle \citep{rosset} at both 100 and 143\,GHz.  The analysis of the \Planck\ data confirms  this upper limit of $1^\circ$ (see Fig.~20 in \cite{delouis2018}). 

\section{Frequency dependence}
\label{sec:moments_expansion}
In this section, we discuss the frequency dependence of the amplitudes of the residuals power spectra plotted in Fig.~\ref{fig:sed_variationp}.

We follow earlier studies \citep{chluba2017, Desert22} using a Taylor expansion of the MBB emission law to model the residuals. 
The moment expansion has been applied to dust power spectra in total intensity by \cite{mangilli2021} and to dust \bmodes~power spectra considered as an intensity by \cite{Azzoni2021} and \citet{vacher2021}.
Within this framework, 
the power spectra of the residual maps is modelled using the first order expansion as
\begin{align}
 \label{eq:moments_definition}
   \mathcal{D}_\ell(\nu) &= \gamma_{\rm P}(\nu)^2 \cdot \bigg\{ \nonumber \\[-0.5mm]
    1^{\rm st}\ \text{order}\ \beta\; &
    \begin{cases}
    &+ \mathcal{D}_\ell^{\omega^{\beta}_1 \times \omega^{\beta}_1} \ln\left(\frac{\nu}{\nu_0}\right)^2 \nonumber \\  \end{cases}\\[-0.5mm]
    1^{\rm st}\ \text{order}\ T \;&
    \begin{cases}
    &+ \mathcal{D}_\ell^{\omega_1^T \times \omega_1^T}\left( \Theta_\nu(T_{\rm d})-\Theta_{\nu_0}(T_{\rm d})\right)^2
    \end{cases}\\[-0.5mm]
    1^{\rm st}\ \text{order}\ T\mathrm{x}\beta \;&
    \begin{cases}
    &+ 2\mathcal{D}_\ell^{\omega^{\beta}_1 \times \omega_1^T}\ln{\left(\frac{\nu}{\nu_0}\right) \cdot \left(\Theta_\nu(T_{\rm d})-\Theta_{\nu_0}(T_{\rm d})\right)}  \bigg\}, \nonumber
     \end{cases}\\[-0.5mm]
     \nonumber
\end{align}
where  $\mathcal{D}_\ell^{\rm a\times b}$ are three moment coefficients introduced by \cite{mangilli2021} and \citet{vacher2021}, which are associated with spatial variations of the dust spectral index and temperature, and correlated variations of these two MBB parameters. The $\Theta_\nu(T_{\rm d})$ function is the derivative of the logarithm of the black-body spectrum with respect to temperature $T_{\rm d}$: 
\begin{eqnarray}
    \label{delta_T}
    \Theta_\nu(T_{\rm d}) = \frac{x}{T_{\rm d}}\frac{e^{x}}{e^{x}-1} \,,
\end{eqnarray}
where $x=\frac{h\nu}{k_BT_{\rm d}}$. The temperature $T_{\rm d}$ is the one used to normalize $\gamma_{\rm P}(\nu)$ in Fig.~\ref{fig:gammap}.

\citet{Ichiki19} and more extensively \citet{Vacher22} have extended the moment expansion to polarization to model the frequency dependence of Stokes $Q$ and $U$ maps. In the formalism introduced by \citet{Vacher22}, the moments are spin-2 objects that characterize the frequency dependence of both polarized intensity and polarization angle. This formalism has not yet been applied to $EE$ and $BB$ power spectra. Qualitatively, one expects that variations of polarization angles 
induce an exchange of power between $E$- and $B$-modes, which is not symmetric due to the E/B asymmetry of dust polarization. The frequency dependence of $EE$ and $BB$ power spectra of residuals maps are thus coupled but not necessarily identical.        
\begin{figure*}[h!]
\centering
  \centering
  \includegraphics[width=1\textwidth]{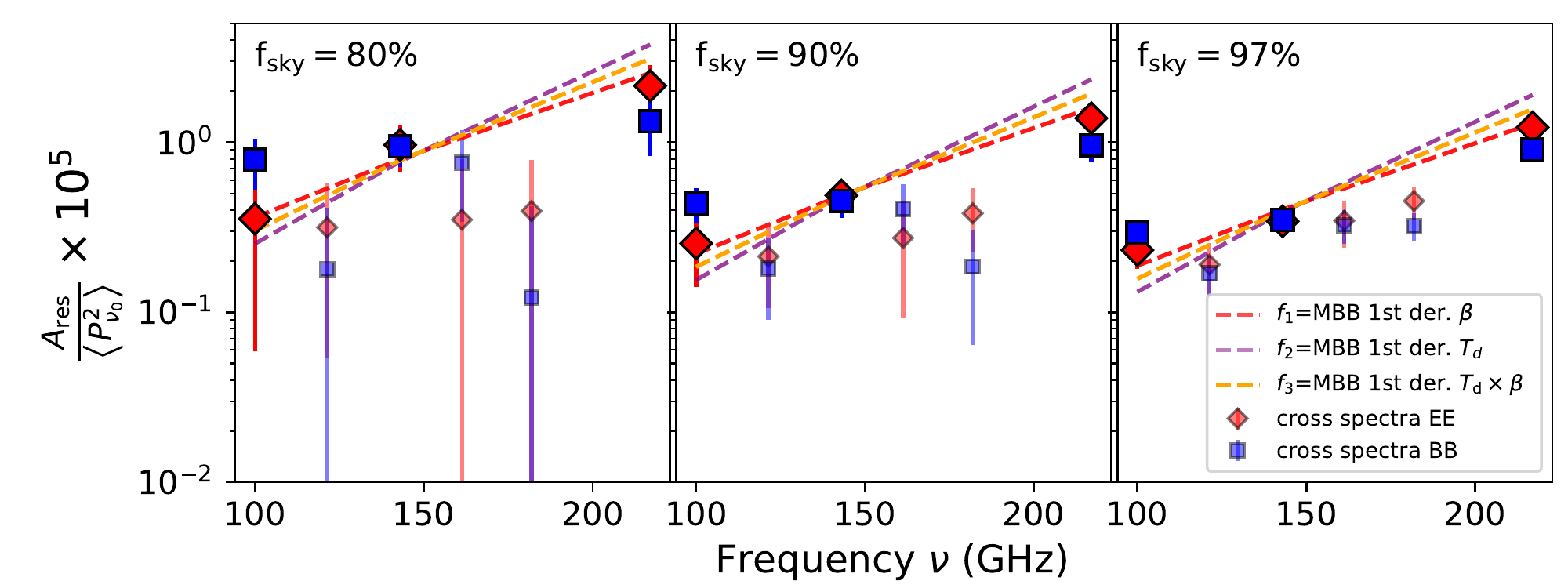}
  \caption{From left to right residual amplitudes $A_{\rm res}(\nu)$ (dark markers) $EE$ (red)\,and $BB$ (blue) for $f_{\rm sky}$=[80,90,97]\%, are shown. A fit to the data $EE$ accounting for the three components of the moment expansion modelling, separately, described in Eq.~\ref{eq:moments_definition} is shown in dashed line. Cross power spectra correlation between pair of frequencies: 100$\times$143, 100$\times$217\,and 143$\times$217 in smaller and lightened markers, is shown.}
  \label{fig:sed_fit}
\end{figure*}
In Fig.~\ref{fig:sed_fit}, we present fits of the  $A_{\rm res}^{EE}(\nu)$
with the three terms in Eq.~\ref{eq:moments_definition}, which we refer to $f_1$, $f_2$ and $f_3$ for $\beta$, $T$ and $T\mathrm{x}\beta$, respectively. 
Over the frequency range of our data analysis from 100 to 217\, GHz, these three functions differ only slightly. The $f_1$ function provides the best fit, especially for $f_{\rm sky} = 97\,\%$, but the fits do not allow us to disentangle the contributions from variations of $\beta_{\rm d}$, $T_{\rm d}$ and their correlation to the power-spectra of the residual maps.  

It is satisfactory to find that the moments expansion to the first  order provides a good model for the $EE$ power-spectra of residual maps. This result suggests that the residuals follow mainly from variations of dust spectral parameters. However, one can see in the Figure that the fits would not be as good for the $BB$ amplitudes, in particular for $f_{\rm sky} = 80\%$ where the amplitudes do not increase with increasing frequency. 

Fig.~\ref{fig:sed_fit} includes amplitudes derived from the 100$\times$143, 100$\times$217\,and 143$\times$217 cross-spectra that are plotted with smaller symbols.
These data points are not used in the fit because their values depend on frequency decorrelation. We do find that the residual maps at the three frequencies are not fully correlated. The 143$\times$217 and to a lesser extent the 100$\times$217 amplitudes lie under the model fit. Part of this mismatch could result from a miscalibration of the absolute polarization angle at 217 GHz. 
To investigate this possibility we performed the following test. We rotated the reference frame of the $Q^{\prime}_{\rm P}$, and $U^{\prime}_{\rm P}$ maps at 217 GHz by an angle $\psi$=$\pm$1$^{\circ}$, which is an upper limit based on the $Planck$ polarization absolute angle uncertainty \citep{delouis2018}. We repeated our analysis with these rotated Stokes maps at 217\,GHz for both the \Planck\ data and the simulations to correct for the impact of the rotation on the CMB signal. We found that the amplitudes of 100$\times$217 and 143$\times$217 do not change coherently for neither $EE$ and $BB$, nor for the three masks considered. 
We conclude from this test that the mismatch between the spectral model and the cross-spectra amplitudes cannot be explained by a miscalibration of the polarization angle in the $Planck$ 217 GHz data. 

\section{Conclusions}
\label{sec:conclusions}
Power spectra of the \Planck\ data are used to characterize  spatial variations of the polarized dust SED. We improve the sensitivity of previous studies by using the newly released \srolltwo\ maps and extending the analysis to regions near the Galactic plane. Our analysis focuses on the lowest multipoles between $\ell$=4 and 32, and three sky areas with $f_{\rm sky} = 80$\%, 90\%, and 97\%. Maps of MBB parameters from the {\it Commander} and GNILC component separation methods applied to the \Planck\ total intensity data are used as reference models. The main results of our analysis are as follows. 

\begin{itemize}
\item 
We confirm earlier studies finding that the mean SED for dust polarization from 100 to 353\,GHz is very close to that for total intensity, and to a MBB spectrum with a spectral index $\beta_{\rm d}$=1.53 for an assumed dust temperature of $T_{\rm d}$ = 19.6K. 
\item 
The mean SED and the 353\,GHz $Q$ and $U$ maps are used to compute residual maps at 100, 143 and 217\,GHz, which quantify spatial variations of the dust polarization SED. Residuals are detected at the three frequencies for the three sky areas. The $EE$ and $BB$ spectra of the residual maps are of comparable amplitude. They do not reproduce the $E/B$ asymmetry observed for the total dust power.

\item
The residual maps are correlated with the reference {\it Commander} and GNILC  models, but this correlation accounts for only a fraction of the residuals amplitude. Further, we find that this fraction decreases toward high Galactic latitudes (i.e. for a decreasing $f_{\rm sky}$). This result shows that models based on total intensity data are underestimating the complexity of dust polarized CMB foreground. Possibly, future developments of {\it Commander} and GNILC models, also accounting for the latest releases of the \Planck~data, could improve the comparison with polarized data.

\item 
To gain insight on the origin of SED variations, we quantify variations in the polarization angle. For 
$f_{\rm sky} =80$\% and 90\%, we find that the contribution of polarization angles to the residuals is dominant, in particular for the $BB$ signal. These results emphasizes the 
importance to consider the geometrical properties of Galactic polarization in component separation.

\item 
The frequency dependence of the $EE$ and $BB$ residual amplitudes yields further insights. We find that the moments expansion to the first order of the MBB spectrum provides a good fit to the $EE$ amplitudes. 
This result suggests that the residuals follow mainly  
from variations of dust spectral parameters (temperature and spectral index). 
However, this conclusion is challenged by the $BB$ results, in particular for $f_{\rm sky} =80$\%, and by cross-spectra that show that the residuals maps at the three frequencies are not fully correlated. 
Further work is needed to model theoretically the impact of polarization angle variations on $EE$ and $BB$ power spectra of residual maps, which we expect to depend on the correlation between dust emission properties and the structure of the magnetized interstellar medium.

\end{itemize}
Our analysis of \Planck\ data brings out significant differences between dust polarization $EE$ and $BB$ SEDs and with respect to total intensity, setting new requirements for simulations of the dust polarized foreground and component separation methods. A significant refinement to dust modelling is necessary to ensure unbiased detection of CMB primordial \bmodes~at the precision required by future CMB experiments.
\begin{acknowledgements}
A.R. acknowledges financial support from the French space agency (Centre National d'Etudes Spatiales, CNES) and the Italian Ministry of University and Research - Project Proposal CIR01\_00010. F.B. acknowledges support from the Agence Nationale
de la Recherche (project BxB: ANR-17-CE31-0022) and CNES. The authors would like to thank the anonymous referee for a thorough reading of the article and for all suggestions and comments that significantly improved the understanding of the text.
\end{acknowledgements}
\bibliographystyle{aat}
\bibliography{biblio}

\end{document}